# Bias-corrected climate projections from Coupled Model Intercomparison Project-6 (CMIP6) for South Asia

Vimal Mishra[1,2], Udit Bhatia[1], Amar Deep Tiwari[1]
1--Civil Engineering, Indian Institute of Technology (IIT) Gandhinagar
2—Earth Sciences, Indian Institute of Technology Gandhinagar
*Corresponding Author: vmishra@iitgn.ac.in

## Abstract

Climate change is likely to pose enormous challenges for agriculture, water resources, infrastructure, and livelihood of millions of people living in South Asia. Here, we develop daily bias-corrected data of precipitation, maximum and minimum temperatures at 0.25° spatial resolution for South Asia (India, Pakistan, Bangladesh, Nepal, Bhutan, and Sri Lanka) and 18 river basins located in the Indian sub-continent. The bias-corrected dataset is developed using Empirical Quantile Mapping (EQM) for the historic (1951-2014) and projected (2015-2100) climate for the four scenarios (SSP126, SSP245, SSP370, SSP585) using output from 13 CMIP6-GCMs. The bias-corrected dataset was evaluated against the observations for both mean and extremes of precipitation, maximum and minimum temperatures. Bias corrected projections from 13 CMIP6-GCMs project a warmer (3-5°C) and wetter (13-30%) climate in South Asia in the 21st century. The bias-corrected projections from CMIP6-GCMs can be used for climate change impact assessment in South Asia and hydrologic impact assessment in the sub-continental river basins.

## Background and Summary

South Asia is one of the most densely populated regions of the world. A majority of the population in South Asia depends on agriculture for their livelihood. South Asia is among the global hot spots that are likely to face the detrimental impacts of climate change[1,2]. Considerable changes in precipitation and temperature are projected in South Asia that will have implications for water resources and agriculture[3–6]. The risk of floods and droughts are likely to increase in South Asia under the warming climate[7–12]. Both recent droughts and floods have affected a large population and caused enormous damage to agriculture and infrastructure in South Asia[13–16]. Similarly, the frequency and intensity of severe heatwaves have increased in South Asia and projected to increase in the future[17–21]. Overall, the frequency of both precipitation and temperature extremes has considerably increased in the past decades and likely to rise further under the warming climate[18,22].

Projections from the General Circulation Models (GCMs) play a vital role in understanding the future changes in climate. However, spatial resolution at which GCMs are run is often too coarse to get reliable projections at the regional and local scale[23]. Precipitation and temperature projections at higher spatial resolution are required for the climate impact assessments[24–26].

Moreover, precipitation and temperature from the GCMs have a bias due to their coarse resolution or model parameterizations[27,28]. Therefore, for the assessment of the climate change and its impacts on different sectors (e.g., water resources, agriculture), bias-correction is required[23,29–34]. Both statistical and dynamical approaches are used for downscaling and bias correction of climate change projections from GCMs. Statistical approaches are based on the distribution and relationship between the observed and projected data for the historical period[33,34]. On the other hand, dynamical downscaling approaches are based on regional climate model forced with the boundary conditions from the coarse resolution GCMs[35,36]. Both statistical and dynamical downscaling approaches have limitations[37,38]. The primary limitation of the dynamical downscaling is related to the requirement of computational efforts to run the regional climate models at higher spatial and temporal resolution[27,39]. Moreover, dynamical downscaling may not remove the bias in climate variables, which might require corrections based on the statistical approaches[39]. Given these limitations, statistical bias correction approaches are widely used in climate change impact assessments[40,41].

Considering the climate change impacts in South Asia, we develop a bias-corrected dataset of daily precipitation, maximum and minimum temperatures using output from 13 GCMs that participated in the Coupled Model Intercomparison Project-6 (CMIP6). The 13 GCMs were selected based on the availability of daily precipitation, maximum and minimum temperatures for the historical and four scenarios (SSP126, SSP245, SSP370, SSP585). We used empirical quantile mapping (EQM) to develop bias-corrected data at daily temporal and 0.25° spatial resolution for six countries in South Asia (India, Pakistan, Bangladesh, Nepal, Bhutan, and Sri Lanka). Also, the bias-corrected projections are developed for 18 sub-continental river basins. The bias-corrected projections from 13 CMIP-GCMs can be used for estimating the projected changes in mean and extreme climate in South Asia. Bias corrected data for 18 sub-continental river basins can be used to develop hydrologic projections using hydrological models.

**Methods**

Bias-corrected projections were developed for South Asia (India, Pakistan, Bangladesh, Nepal, Bhutan, and Sri Lanka) and the 18 Indian sub-continental river basins (Figure 1). We used basin boundaries in the Indian sub-continent from Shah and Mishra (2016)[42] [Fig. 1]. We obtained observed daily gridded precipitation, minimum and maximum temperatures for South Asia for the 1951-2018 period. Daily precipitation at 0.25° was obtained from the India Meteorological Department (IMD) for the Indian region[43]. Pai et al. (2014)[43] developed gridded daily precipitation for India using station observations from more than 6000 stations located across India. The precipitation captures critical features of the Indian summer monsoon, including higher rainfall in the Western Ghats and northeastern India and lower rainfall in the semi-arid and arid regions of western India. Besides, gridded precipitation captures the orographic rain in the Western Ghats and foothills of Himalaya. The gridded precipitation data from IMD has been used for various hydroclimatic applications[13,44,45]. Gridded daily maximum and minimum temperatures from IMD were developed using station-based observations from more than 350 stations located across India[46]. There is bias in temperature observations from IMD in the Himalayan region, which can be attributed to sparse station density[44,47]. Gridded precipitation and maximum and minimum temperatures were obtained from Sheffield et al. (2006)[48] for the regions outside India. Datasets from Sheffield et al. (2006)[48] are available at 0.25° spatial and daily temporal resolutions. Consistency between IMD and Sheffield et al. (2006)[48] dataset was checked in Shah and Mishra (2016)[42], who reported that Sheffield et al. (2006)[48] dataset has a good agreement with the IMD observations. Nonetheless, we used IMD gridded dataset for the Indian region and Sheffield et al. (2006)[48] for outside India for bias correction of projections from CMIP6 as the IMD data is widely used for hydroclimatic studies in India.

We obtained daily precipitation, maximum and minimum temperatures from 13 CMIP6-GCMs from https://esgf-node.llnl.gov/search/cmip6/. All the three variables and for all the scenarios were available only for these 13 GCMs. Therefore, we restricted bias-correction to only these models. Precipitation, maximum and minimum temperatures from CMIP6-GCMs are available at different spatial resolutions (Table S1). For instance, the spatial resolution of the CMIP6 projection varies from 0.7° (EC-Earth3) to more than 2° (CanESM5). All the three variables were selected for the historical (1850-2014), ssp126 (2015-2100), ssp245 (2015-2100), ssp370 (2015-2100), and ssp585 (2015-2100) scenarios under r1i1p1f1 initial condition at daily time scale[49]. The scenarios used in the CMIP6 combine Shared Socioeconomic Pathways (SSP) and target radiative forcing levels at the end of the 21$^{st}$ century[50]. For instance, SSP126 indicates SSP-1 and target radiative forcing at the end of the 21$^{st}$ century 2.6 Watt/m$^2$. Therefore, SSP126 is a mitigation scenario. On the other hand, SSP585 is based on the emission scenario considering SSP-5 and radiative forcing of 8.5 Watt/m$^2$ at the end of the 21$^{st}$ century[50]. Further details on the scenarios used in the CMIP6 can be obtained from Gidden et al. (2019)[50]. We regridded all the variables from CMIP6 to 1° spatial resolution to make them consistent. However, the effect of regridding using bilinear interpolation was checked by comparing the gridded datasets against the raw data for all-India mean of precipitation, maximum and minimum temperatures. We did not find any considerable differences in the all-India averaged precipitation and temperature using regridded and raw output from the GCMs.

Outputs of the various atmospheric (e.g., maximum and minimum temperatures, and precipitation) variables obtained from GCMs are known to exhibit systematic biases (Fig. S1). Hence, these outputs need to be bias-corrected to produce reliable estimates at regional and local scales for climate impact assessment. To achieve this, statistical transformations that attempt to find a function that maps the model output to a new distribution such that the resulting distribution matches that of observations. In general, this transformation can be formulated as Piani et al. (2010)[30]:

$$x_m^o = f(x_m) \quad (1)$$

Where $x_m^o$ is the bias-corrected model output. If the statistical distribution of $x_m$ and $x_0$ are known, the transformation can be written as:

$$x_m^o = F_0^{-1}(F_m(x_m)) \quad (2)$$

Where $F_m$ and $F_o$ are the Cumulative Distribution Functions (CDFs) of $x_m$ and $x_o$ respectively. In Empirical Quantile Mapping (EQM[51,52]), instead of assuming parametric distributions, empirical CDFs[34,53,54] are estimated from the percentiles calculated from $x_m$ and $x_0$. As a result, EQM and its variants can be applied to both temperature and precipitation even if their underlying distributions are different and hence recommended for statistical bias correction[55].

In the context of statistical downscaling, since the observations are at a higher resolution than models, EQM on bilinearly interpolated model outputs at observation resolution is often used to address the scale mismatch and generate post-processed model outputs[52]. We choose non-parametric transformation approaches over the parametric approaches as has shown better skills in the comparison to parametric methods in reducing biases from GCM as well as Regional Climate Model (RCM) outputs[56].

We used EQM to statistically downscale the daily maximum and minimum temperatures, and precipitation for South Asia and Indian sub-continental river basins (Fig. 1). We use the outputs ($x_m$) from 13 CMIP6-GCMs (Table S1), which are available at different resolutions (Table S1). Observations for the three variables at the resolution of 0.25-degree are obtained from the IMD, Pai et al. (2014)[43] for Indian Region and Sheffield et al. (2006)[48] for grid-points within and outside India, respectively. We used the 1951-2014 period to obtain the transformation function to map the distribution of $x_m$ to $x_o$. For precipitation, the drizzle effect is corrected by using a wet day threshold of 1 mm/day[30,56]. If the values from model projections are larger (smaller) than the training values used to estimate the empirical CDF, the correction found for the highest (lowest) quantile of the training period is used. We used mapped transformation to bias correct the outputs for the historical period and the SSP126, SSP245, SSP370, and SSP585 scenarios for the 2015-2100 period for all the three variables. Raw and bias-corrected data for INM-CM5 is shown against the observed maximum temperature for a randomly selected grid in the Indian subcontinent (Fig. S1). Quantile mapping based statistical bias correction has been widely used, and its performance was found to be satisfactory in comparison to the other methods[25,34,57].

**Data Record**

Bias corrected daily precipitation, maximum and minimum temperatures are available for the 13 GCMs (Table S1) for the historical (1951-2014) and future (2015-2100) periods. Projections for the future are available for the four scenarios (SSP126, SSP245, SSP370, and SSP585) for South India (India, Pakistan, Bangladesh, Sri Lanka, Bhutan, and Nepal) and Indian sub-continental river basins (Fig. 1). The basin wise dataset (Data Citation 1) and country wise dataset (Data Citation 2) has been made available through Zenodo. Details on the data format can be obtained from a readme file provided at the above link.

**Technical Validation**

First, we estimated the projected changes in mean annual precipitation, maximum and minimum temperatures using the raw data from the CMIP6-GCMs (Fig. 2). The projected changes were estimated for each GCM for the late 21$^{st}$ century (2074-2100) against the historical reference period (1988-2014). Then, the multimodel ensemble mean of the projected changes from all the 13 CMIP6-GCMs was taken. The multimodel ensemble mean annual precipitation is projected to increase in South Asia under the projected future climate (Fig. 2a-d). The projected increase in precipitation in South Asia under the future climate varies with the scenario considered. For instance, under the high-emission scenario (SSP585), a considerably higher increase (more than 30%) in the multimodel ensemble mean is projected in comparison to the low emission (SSP126) scenario (less than 13%). Similarly, there are regional differences in the projections of precipitation from the CMIP6-GCMs. For example, a more substantial increase in the multimodel ensemble mean precipitation is projected for the semi-arid and arid regions of western South Asia than the other regions (Fig. 1a-d). Similar to rainfall, mean annual maximum and minimum temperatures are projected to rise substantially in South Asia under the future climate (Fig. 2). Projected changes in mean annual minimum temperature are generally greater than the changes in mean annual maximum temperature. As expected, the high emission scenario (SSP585) will lead to a much higher rise in temperatures than the low emission scenario of SSP126 (Fig. 2).

The raw datasets of precipitation, maximum and minimum temperatures can be used to estimate the projected changes in South Asia under the future climate for different scenarios. However, climate impact studies need bias-corrected projections for decision making at regional and local scales. Since the bias-corrected dataset is consistent with observation for a climatological mean period, it is easier to infer the project changes and its implications in different sectors (e.g., water resources and agriculture) for observations. We, therefore, bias-corrected precipitation, maximum and minimum temperatures for the historical (1951-2014) and future (2015-2100) periods for all the four scenarios for South Asia and Indian sub-continental river basins. The bias-corrected dataset can be used for any region or river basin in South Asia or the Indian sub-continent (Fig. 1).

We estimated the multimodel ensemble mean bias in precipitation, maximum and minimum temperatures from the 13 CMIP6-GCMs (Fig. 3). The bias in mean annual precipitation, maximum and minimum temperatures was estimated against the observations from IMD (for the Indian domain) and Sheffield et al. (2006)[48] observations (for outside India). The CMIP6-GCMs show a dry bias (15-20%) in mean annual precipitation in the majority of South Asia (Fig. 1a). On the other hand, the multimodel ensemble mean positive bias in mean annual precipitation was found in the regions located in Nepal, Pakistan, and Peninsular India (Fig. 3a). A high cold bias in both mean annual maximum and minimum temperatures were found in the Himalayan region in the CMIP6-GCMs (Fig. 3c,e). Also, CMIP6-GCMs exhibit warm bias in mean annual minimum temperature in the majority of South Asia except for the Himalayan region (Fig. 3e). We applied the EQM approach to correct the bias in the CMIP6-GCM output at daily timescale. The bias was substantially reduced after the bias correction in all the three variables for the historical (1985-2014) period (Fig. 3b-f). The reduction in bias in mean annual precipitation, maximum and minimum temperatures shows the effectiveness of our bias correction approach based on EQM.

Similar to mean annual precipitation, maximum and minimum temperatures, we estimated bias in precipitation and temperatures extremes in the raw output from the CMIP6-GCMs (Fig. 4). The 90$^{th}$ percentiles of precipitation of rainy days (precipitation more than 1mm), maximum and minimum temperatures were compared for the historical period (1985-2014) from CMIP6-GCMs against the observed dataset. Consistent with mean annual precipitation, a considerable dry bias is present in extreme precipitation in CMIP6-GCMs across South Asia (Fig. 4a). We find that the CMIP6-GCMs show a warm bias in the 90$^{th}$ percentile of maximum and minimum temperatures across South Asia except in the Himalayan region (Fig. 4c, e). In the Himalayan region, a cool bias in CMIP6-GCMs in maximum and minimum temperature extreme was found (Fig. 4c, e). We find that the EQM based bias correction has successfully removed the bias in extreme precipitation, maximum and minimum temperatures across South Asia (Fig. 4). Therefore, the bias in both mean and extremes of precipitation, maximum and minimum temperatures were removed. Also, we compared the season cycle of bias-corrected precipitation, maximum and minimum temperatures from the CMIP6-GCMs against the observed dataset for the 1985-2014 period. Uncertainty in the bias-corrected precipitation, maximum and minimum temperatures were estimated using one standard deviation. We find that the seasonal cycle of the multimodel ensemble mean bias-corrected precipitation, maximum, and minimum temperatures compare well against the observations (Fig. 5). Overall, our results show that the EQM approach successfully corrects the bias in the CMIP6-GCMs, which can be used for climate impacts studies in South Asia. Also, the bias-corrected dataset can be used for hydrological studies in the Indian sub-continental river basins.

**Climate Projections for South Asia and Indian sub-continental River basins**

Daily bias-corrected projections of precipitation, maximum and minimum temperatures at 0.25° from CMIP6-GCMs are developed for South Asia and the 18 Indian sub-continental river basins (Fig. 1). The projections are available for the historical (1951-2014) and future (2015-2100) periods. We estimated projected changes in precipitation, maximum and minimum temperatures in the late 21$^{st}$ century from the bias-corrected dataset against the historical reference (1988-2014) period for all the scenarios (SSP126, SSP245, SSP370, and SSP585) [Fig. 6]. Our bias-corrected precipitation projections show consistent spatial patterns that were observed in the raw CMIP6-GCMs (Fig. 2). For instance, a larger increase in mean annual precipitation was found in the western parts of South Asia in both raw and bias-corrected datasets (Fig. 6a-d). A considerably large increase in mean annual precipitation is projected under SSP585 (median 23%) than under SSP126 (median 12%) [Fig. 6a-d]. The ensemble mean median change in maximum temperature is projected to be 1.3°C in SSP126 and 2.2°C in SSP585 (Fig. 6e-h). Similarly, the ensemble mean minimum temperature is projected to rise significantly across South Asia with a median increase of more than 3°C increase in SSP585 scenario (Fig. 6i-l).

Projected changes in precipitation, maximum and minimum temperatures were estimated using a 30-year moving window for all the six countries in South Asia under the highest emission scenario of SSP585 (Fig. 7). We considered the SSP585 scenario to estimate the projected change in precipitation, maximum and minimum temperatures under the worst case (Fig. 7). Projected change in each CMIP6-GCM was estimated for each 30-year window (1986-2015, 1987-2016 … 2071-2100) against the historical reference period of 1985-2014. Moreover, we estimated uncertainty in the bias-corrected CMIP6-GCMs using one standard deviation of projected change in the individual GCMs. The multimodel ensemble mean annual precipitation is projected to rise in all the six countries under the future climate (Fig. 7). All the six countries in South Asia are projected to experience a 20-40% rise in mean annual precipitation under the SSP585 scenario by the end of the 21$^{st}$ century. However, the bias-corrected precipitation projections show more uncertainty for Pakistan than the other countries (Fig. 7). Uncertainty in the bias-corrected maximum and minimum temperatures is substantially lesser than that of precipitation (Fig. 7). The multimodel ensemble mean bias-corrected mean annual maximum temperature is projected to rise by 3-4°C by the end of the 21$^{st}$ century under SSP585. Moreover, the bias-corrected ensemble mean annual minimum temperature is projected to rise by 3-5°C by the end of the 21$^{st}$ century (Fig. 7). We find a different level of uncertainty in mean annual precipitation, maximum and minimum temperatures for the six countries in South Asia (Fig. 7). Overall, the climate is projected to become wetter and warmer in South Asia in the future, and the magnitude of change will depend on the scenarios.

We estimated projected changes in mean annual precipitation, maximum and minimum temperatures for the six countries, and 18 sub-continental river basins for the Near (2020-2046), Mid (2047-2073), and Far (2074-2100) periods against the historical reference of 1988-2014 (Table S2-S7). The multimodel projected changes were estimated for all the four scenarios along with the mean for the historical period (Table S2-S7). The multimodel ensemble mean bias-corrected precipitation is projected to change between 3-20% in the Near term under the SSP126 (Table S2). The most substantial increase in precipitation is projected in Pakistan, while the lowest rise is expected in Bhutan. Precipitation is projected to rise substantially in the Far term in all the countries in South Asia under the SSP585 (Table S2). The ensemble mean bias-corrected precipitation is projected to change by 31-53%, with the most considerable projected rise in Pakistan in the Far term under SSP585 (Table S2). The projected increase in mean annual maximum temperature is far lesser (0.48-0.97°C) in the Near term under SSP126 in comparison to the Far (2.6-5.3°C) term under the SSP585 (Table S3) in the six countries in South Asia. Our bias-corrected data based on 13 CMIP6-GCMs projected an increase of 0.7-1.3°C) in mean annual minimum temperature in the Near term under SSP126 (Table S4).

Moreover, mean annual minimum temperature is projected to rise between 3.5 to 5.5°C in the late 21$^{st}$ century under the SSP585 (Table S4). Uncertainty in the projections of bias-corrected precipitation, maximum and minimum temperatures was estimated for each country and period under all the four scenarios (Table S2-S4). Temperature projections show lesser uncertainty than the projections of precipitation, which might have implications for hydrologic applications of the bias-corrected projections (Table S2-S4).

We estimated projected changes in mean annual precipitation, maximum and minimum temperatures using bias-corrected data from the 13 CMIP6-GCMs for the 18 sub-continental river basins (Table S5-S7, Fig. 1). Bias corrected projections of the three climatic variables are essential for hydrologic modelling, and the climate change impact assessment. Mean annual precipitation is projected to rise across the basins under all the scenarios in the projected future climate (Table S5). The projected rise in the mean annual precipitation is considerably higher in the SSP585 in comparison to the SSP126. The projected rise in precipitation in the sub-continental basins is higher in the Far period than the Near period.

Notwithstanding a considerable uncertainty in the precipitation projections, bias-corrected data show that precipitation is projected to rise more in the river basins located in the semi-arid/arid regions of the Indian sub-continent (Table S5). Similarly, significant warming in the mean annual maximum and minimum temperatures is projected based on the bias-corrected data from the 13 CMIP6-GCMs (Table S6-S7). Mean annual maximum temperature is projected to rise between 2.5-4.4°C in the Far period under SSP585 in the Indian sub-continental river basins (Table S6). Moreover, the mean annual minimum temperature is projected to rise by 3.0-5.0°C in the Far period under SSP585 in the river basins of the Indian sub-continent (Table S7). Basin specific projections and associated uncertainty can be seen in supplemental Tables S5-S7. Overall, the bias-corrected projections can be used for the hydroclimatic impact assessment in the sub-continental river basins.

**Usage Notes**

Daily bias-corrected projections of precipitation, maximum and minimum temperatures at 0.25° are essential for climate impact assessment for the administrative boundaries or at river basin scale[25]. We developed bias-corrected projections from 13 CMIP6-GCMs that can be used for hydroclimatic impact assessment based on mean and extremes in South Asia. Bias corrected data performs well against the mean and extremes. The dataset has been arranged based on the geographical boundaries of six countries in South Asia. Moreover, we provide separate data for each of the 18 sub-continental river basins. Daily bias-corrected projections can be used to estimate climatic indices associated to mean and extremes. For instance, daily maximum and minimum temperatures can be used to estimated projected changes under different scenarios for the crop growing seasons. Moreover, the temperature dataset can be used to estimate growing degree days (GDD)[58] and other indicators of extreme heat during the crop growing period[59].

Similarly, daily precipitation projections can be used to estimate changes in mean and extreme precipitation for any period during the 21$^{st}$ century[27,60]. Data users can also estimate the differences in indicators and potential impacts based on the low (SSP126) and high (SSP585) emission scenarios. Most of the hydrological models require daily precipitation, maximum and minimum temperatures as the primary inputs of meteorological forcing. Therefore, hydrological models can be used with the bias-corrected projections to estimate the impacts of the projected future climate on hydrology for a river basin or a region.

As an example, we use the bias-corrected projections to estimate the frequency of precipitation and temperature extremes for an administrative region (state of Uttar Pradesh, India) and a river basin (Godavari, India) [Figure 8-9]. The frequency of extreme precipitation was estimated using 95$^{th}$ percentiles of rainy days (precipitation more than 1mm). Similarly, the frequency of extreme hot maximum and minimum temperatures was estimated using the 95$^{th}$ percentile of the two hottest months (April-May) in the region. As expected, both precipitation and temperature extremes are projected to rise in Uttar Pradesh and Godavari basin under the SSP585 scenario (Fig 8-9). The projected rise in the frequency of precipitation and temperature extremes is higher for the Far period than the Near-term climate. Overall, daily bias-corrected CMIP6 projections can be used for multiple assessments related to climate and hydrology in one of the most populated regions of the world. In the future, bias-corrected projections will be made available from more CMIP6-GCMs as their output becomes available. More details on the data can be found in the link and from the readme file.

**Acknowledgement:** We acknowledge the data availability from India Meteorology Department (IMD) and Sheffield et al. (2006). Output from the CMIP6 models from https://esgf-node.llnl.gov/projects/cmip6/ is greatly acknowledged.

**Code Availability:** Codes used for bias correction of CMIP6-GCMs can be made available on request.

**Author Contributions:** VM designed the study and wrote the manuscript. AT downloaded and processed CMIP6 projections. UB did the bias correction of CMIP6 projections.

**Competing Interests:** Authors declare no competing interests.

References

1. Suarez-Gutierrez, L., Müller, W. A., Li, C. & Marotzke, J. Dynamical and thermodynamical drivers of variability in European summer heat extremes. *Clim. Dyn.* **54**, 4351–4366 (2020).
2. Diffenbaugh, N. S. & Giorgi, F. Climate change hotspots in the CMIP5 global climate model ensemble. *Clim. Change* **114**, 813–822 (2012).
3. Knox, J., Hess, T., Daccache, A. & Wheeler, T. Climate change impacts on crop productivity in Africa and South Asia. *Environ. Res. Lett.* **7**, (2012).
4. Turner, A. G. & Annamalai, H. Climate change and the South Asian summer monsoon. *Nat. Clim. Chang.* **2**, 587–595 (2012).
5. Lobell, D. B. & Burke, M. B. Why are agricultural impacts of climate change so uncertain? the importance of temperature relative to precipitation. *Environ. Res. Lett.* **3**, (2008).
6. Immerzeel, W. W., van Beek, L. P. H. & Bierkens, M. F. P. Climate change will affect the Asian water towers. *Science* **328**, 1382–5 (2010).
7. Aadhar, S. & Mishra, V. A substantial rise in the area and population affected by dryness in South Asia under 1.5 °c, 2.0 °c and 2.5 °c warmer worlds. *Environ. Res. Lett.* **14**, 0–12 (2019).
8. Aadhar, S. & Mishra, V. Increased drought risk in South Asia under warming climate: Implications of uncertainty in potential evapotranspiration estimates. *J. Hydrometeorol.* 1–46 (2020) doi:10.1175/jhm-d-19-0224.1.
9. Ali, H., Modi, P. & Mishra, V. Increased flood risk in Indian sub-continent under the warming climate. *Weather Clim. Extrem.* **25**, 100212 (2019).
10. Hirabayashi, Y. *et al.* Global flood risk under climate change. *Nat. Clim. Chang.* **3**, 816–821 (2013).
11. Arnell, N. W. & Gosling, S. N. The impacts of climate change on river flood risk at the global scale. *Clim. Change* **134**, 387–401 (2016).
12. Winsemius, H. C. *et al.* Global drivers of future river flood risk. *Nat. Clim. Chang.* **6**, 381–385 (2016).
13. Mishra, V., Aadhar, S., Asoka, A., Pai, S. & Kumar, R. On the frequency of the 2015 monsoon season drought in the Indo-Gangetic Plain. *Geophys. Res. Lett.* **43**, 12,102-12,112 (2016).
14. Webster, P. J., Toma, V. E. & Kim, H. M. Were the 2010 Pakistan floods predictable? *Geophys. Res. Lett.* **38**, 1–5 (2011).
15. Hunt, K. M. R. & Menon, A. The 2018 Kerala floods: a climate change perspective. *Clim. Dyn.* **54**, 2433–2446 (2020).
16. Mishra, V. & Shah, H. L. Hydroclimatological Perspective of the Kerala Flood of 2018. *J.*

*Geol. Soc. India* **92**, 645–650 (2018).
17. Mishra, V., Mukherjee, S., Kumar, R. & Stone, D. A. Heat wave exposure in India in current, 1.5 °c, and 2.0 °c worlds. *Environ. Res. Lett.* **12**, (2017).
18. J S, N. *et al.* A seven-fold rise in the probability of exceeding the observed hottest summer in India in a 2 °C warmer world. *Environ. Res. Lett.* **15**, 044028 (2020).
19. Mukherjee, S. & Mishra, V. A sixfold rise in concurrent day and night-time heatwaves in India under 2 °C warming. *Sci. Rep.* **8**, 1–9 (2018).
20. Im, E. S., Pal, J. S. & Eltahir, E. A. B. Deadly heat waves projected in the densely populated agricultural regions of South Asia. *Sci. Adv.* **3**, 1–8 (2017).
21. Mazdiyasni, O. *et al.* Increasing probability of mortality during Indian heat waves. *Sci. Adv.* **3**, 1–6 (2017).
22. Mukherjee, S., Aadhar, S., Stone, D. & Mishra, V. Increase in extreme precipitation events under anthropogenic warming in India. *Weather Clim. Extrem.* **20**, 45–53 (2018).
23. Christensen, J. H., Boberg, F., Christensen, O. B. & Lucas-Picher, P. On the need for bias correction of regional climate change projections of temperature and precipitation. *Geophys. Res. Lett.* **35**, (2008).
24. Cayan, D. R. *et al.* Future dryness in the Southwest US and the hydrology of the early 21st century drought. *Proc. Natl. Acad. Sci. U. S. A.* **107**, 21271–21276 (2010).
25. Maurer, E. P., Hidalgo, H. G., Das, T., Dettinger, M. D. & Cayan, D. R. The utility of daily large-scale climate data in the assessment of climate change impacts on daily streamflow in California. *Hydrol. Earth Syst. Sci.* **14**, 1125–1138 (2010).
26. Barbero, R., Fowler, H. J., Lenderink, G. & Blenkinsop, S. Is the intensification of precipitation extremes with global warming better detected at hourly than daily resolutions? *Geophys. Res. Lett.* **44**, 974–983 (2017).
27. Mishra, V. *et al.* Reliability of regional and global climate models to simulate precipitation extremes over India. *J. Geophys. Res. Atmos. Res.* **119**, 9301–9323 (2014).
28. Ashfaq, M., Rastogi, D., Mei, R., Touma, D. & Ruby Leung, L. Sources of errors in the simulation of south Asian summer monsoon in the CMIP5 GCMs. *Clim. Dyn.* **49**, 193–223 (2017).
29. Maraun, D. *et al.* Towards process-informed bias correction of climate change simulations. *Nat. Clim. Chang.* **7**, 764–773 (2017).
30. Piani, C., Haerter, J. O. & Coppola, E. Statistical bias correction for daily precipitation in regional climate models over Europe. *Theor. Appl. Climatol.* **99**, 187–192 (2010).
31. Eisner, S., Voss, F. & Kynast, E. Statistical bias correction of global climate projections - Consequences for large scale modeling of flood flows. *Adv. Geosci.* **31**, 75–82 (2012).
32. Wood, A. W., Leung, L. R., Sridhar, V. & Lettenmaier, D. P. Hydrologic implications of dynamical and statistical approaches to downscaling climate model outputs. *Clim. Change* **62**, 189–216 (2004).
33. Pierce, D. W., Cayan, D. R., Maurer, E. P., Abatzoglou, J. T. & Hegewisch, K. C. Improved bias correction techniques for hydrological simulations of climate change. *J. Hydrometeorol.* **16**, 2421–2442 (2015).
34. Thrasher, B., Maurer, E. P., McKellar, C. & Duffy, P. B. Technical Note: Bias correcting climate model simulated daily temperature extremes with quantile mapping. *Hydrol. Earth Syst. Sci.* **16**, 3309–3314 (2012).
35. Giorgi, F. & Gutowski, W. J. Regional Dynamical Downscaling and the CORDEX Initiative. *Annu. Rev. Environ. Resour.* **40**, 467–490 (2015).

36. Mearns, L. O. *et al.* Climate change projections of the North American Regional Climate Change Assessment Program (NARCCAP). *Clim. Change* **120**, 965–975 (2013).
37. Abatzoglou, J. T. & Brown, T. J. A comparison of statistical downscaling methods suited for wildfire applications. *Int. J. Climatol.* **32**, 772–780 (2012).
38. Maurer, E. P. & Hidalgo, H. G. Utility of daily vs. monthly large-scale climate data: an intercomparison of two statistical downscaling methods. *Hydrol. Earth Syst. Sci.* **14**, 1125–1138 (2008).
39. White, R. H. & Toumi, R. The limitations of bias correcting regional climate model inputs. *Geophys. Res. Lett.* **40**, 2907–2912 (2013).
40. Gutmann, E. *et al.* An intercomparison of statistical downscaling methods used for water resource assessments in the United States. *Water Resour. Res. Res.* **50**, 7167–7186 (2014).
41. Xu, L. & Wang, A. Application of the Bias Correction and Spatial Downscaling Algorithm on the Temperature Extremes From CMIP5 Multimodel Ensembles in China. *Earth Sp. Sci.* **6**, 2508–2524 (2019).
42. Shah, H. L. & Mishra, V. Hydrologic Changes in Indian Sub-Continental River Basins (1901-2012). *J. Hydrometeorol.* (2016) doi:10.1175/JHM-D-15-0231.1.
43. Pai, D. S. *et al.* Development of a new high spatial resolution (0.25° × 0.25°) long period (1901-2010) daily gridded rainfall data set over India and its comparison with existing data sets over the region. *Mausam* **65**, 1–18 (2014).
44. Aadhar, S. & Mishra, V. High-resolution near real-time drought monitoring in South Asia. *Sci. Data* **4**, 1–14 (2017).
45. Mishra, V. Long-term (1870–2018) drought reconstruction in context of surface water security in India. *J. Hydrol.* **580**, 124228 (2020).
46. Srivastava, A. K., Rajeevan, M. & Kshirsagar, S. R. Development of a high resolution daily gridded temperature data set ( 1969 – 2005 ) for the Indian region. *Atmos. Sci. Lett.* **10**, 249–254 (2009).
47. Shah, R. & Mishra, V. Evaluation of the Reanalysis Products for the Monsoon Season Droughts in India. *J. Hydrometeorol.* **15**, 1575–1591 (2014).
48. Sheffield, J., Goteti, G. & Wood, E. F. Development of a 50-year high-resolution global dataset of meteorological forcings for land surface modeling. *J. Clim.* **19**, 3088–3111 (2006).
49. Eyring, V. *et al.* Overview of the Coupled Model Intercomparison Project Phase 6 (CMIP6) experimental design and organization. *Geosci. Model Dev.* **9**, 1937–1958 (2016).
50. Gidden, M. J. *et al.* Global emissions pathways under different socioeconomic scenarios for use in CMIP6: A dataset of harmonized emissions trajectories through the end of the century. *Geosci. Model Dev.* **12**, 1443–1475 (2019).
51. Wood, A. W., Maurer, E. P., Kumar, A. & Lettenmaier, D. P. Long-range experimental hydrologic forecasting for the eastern United States. *J. Geophys. Res. D Atmos.* **107**, 1–15 (2002).
52. Aadhar, S. & Mishra, V. High-resolution near real-time drought monitoring in South Asia. *Sci. Data* **4**, 170145 (2017).
53. Julien, B., L., T., F., H. & E., M. Statistical and dynamical downscaling of the Seine basin climate for hydro-meteorological studies. *Int. J. Climatol.* **27**, 1643–1655 (2007).
54. Jakob Themeßl, M., Gobiet, A. & Leuprecht, A. Empirical-statistical downscaling and error correction of daily precipitation from regional climate models. *Int. J. Climatol.* **31**, 1530–1544 (2011).

55. Cannon, A. J. Quantile regression neural networks: Implementation in R and application to precipitation downscaling. *Comput. Geosci.* **37**, 1277–1284 (2011).
56. Gudmundsson, L., Bremnes, J. B., Haugen, J. E. & Engen-Skaugen, T. Technical Note: Downscaling RCM precipitation to the station scale using statistical transformations – A comparison of methods. *Hydrol. Earth Syst. Sci.* **16**, 3383–3390 (2012).
57. Bürger, G., Murdock, T. Q., Werner, A. T., Sobie, S. R. & Cannon, A. J. Downscaling extremes-an intercomparison of multiple statistical methods for present climate. *J. Clim.* **25**, 4366–4388 (2012).
58. Meng, Q. & Mourshed, M. Degree-day based non-domestic building energy analytics and modelling should use building and type specific base temperatures. *Energy Build.* **155**, 260–268 (2017).
59. Ben-Ari, T. *et al.* Identifying indicators for extreme wheat and maize yield losses. *Agric. For. Meteorol.* **220**, 130–140 (2016).
60. Mukherjee, S., Aadhar, S., Stone, D. & Mishra, V. Increase in extreme precipitation events under anthropogenic warming in India. *Weather Clim. Extrem.* **20**, 45–53 (2018).

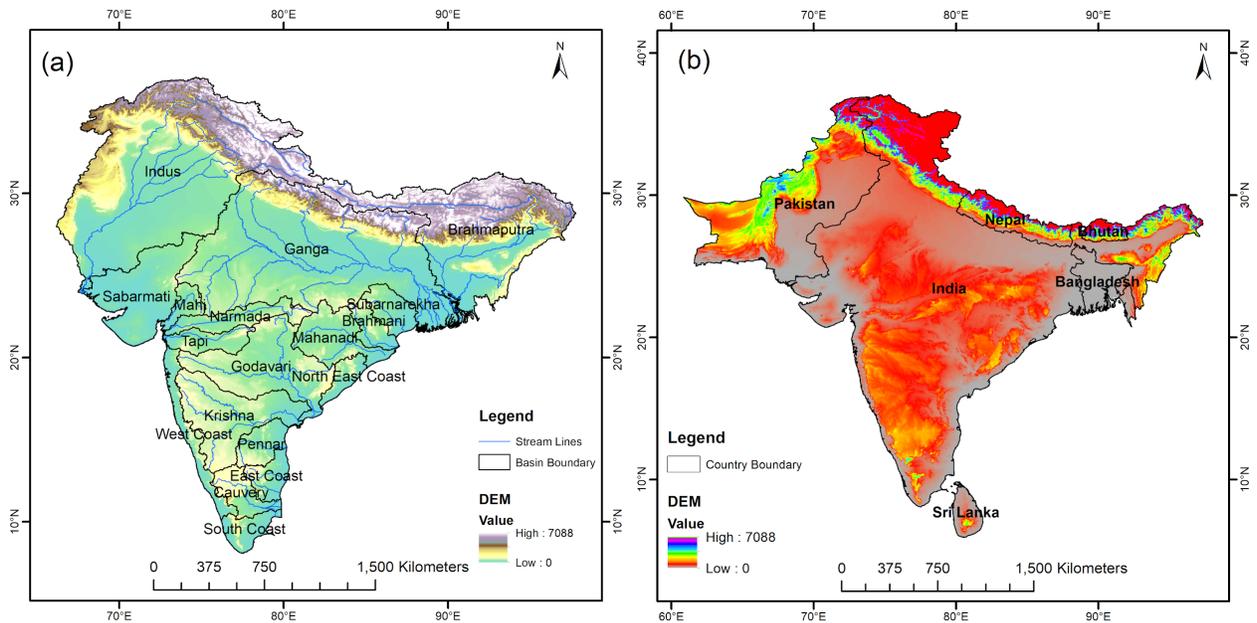

Figure 1. **Geographical domains for bias-corrected CMIP6 projections.** (a) Indian Subcontinent river basin boundaries (black) with the streamlines (blue). Topography in the color scale is shown in the background. Names of the sub-continental river basins are written within the basin boundaries. (b) South Asian country boundaries (black) where topography in the color scale is shown in the background.

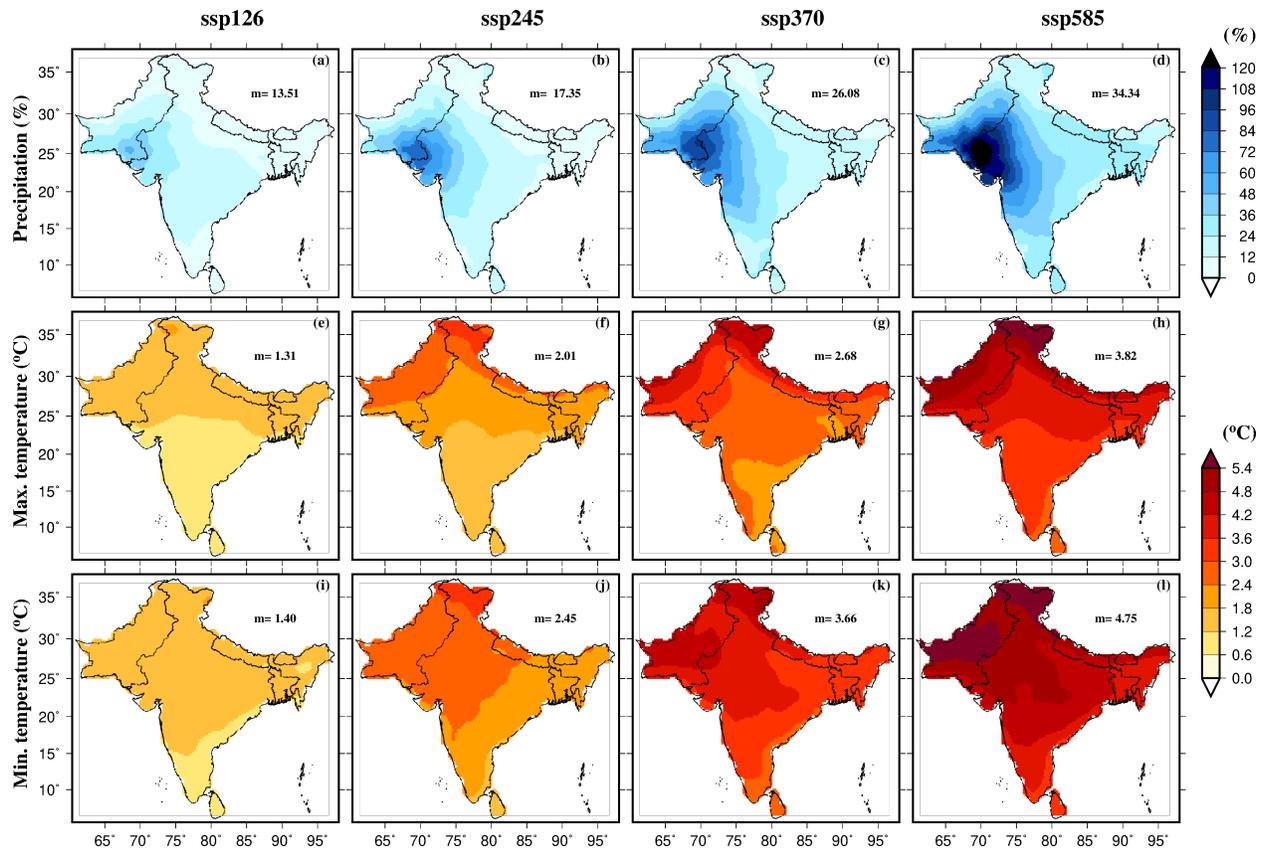

Figure 2. **Projections of precipitation, maximum and minimum temperatures for the end of 21st century using raw output from CMIP6-GCMs.** (a-d) Multimodel ensemble mean projected change in mean annual precipitation (%) for the Far (2074-2100) with respect to the historical period (1988-2014), (e-h) same as (a-d) but for the mean annual maximum temperature, (i-l) same as (a-d) but for the mean annual minimum temperature. Median of the multimodel ensemble mean precipitation; maximum and minimum temperatures are shown in each panel. Projected changes were estimated for the four scenarios (SSP126, SSP245, SSP370, and SSP585) against the historical period.

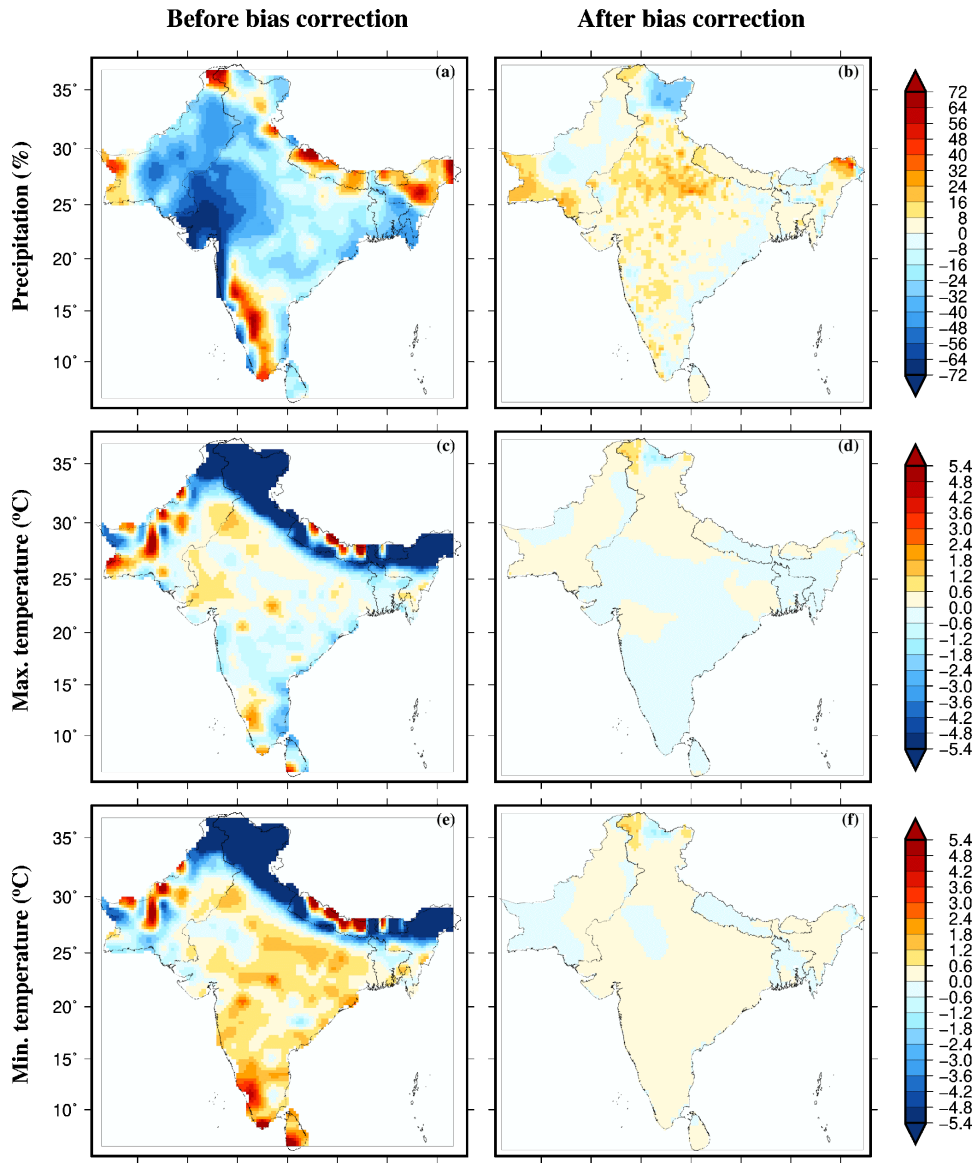

Figure 3. **Multimodel ensemble mean bias in precipitation, maximum and minimum temperatures in 13 CMIP6-GCMs.** (a) Bias (%) in mean annual precipitation for the historical period (1985-2014), (b) bias in mean annual precipitation (%) after the bias correction, (c,d) Bias (°C) in mean annual maximum temperature before and after bias correction, and (e,f) bias (°C) in mean annual minimum temperature before and after bias correction.

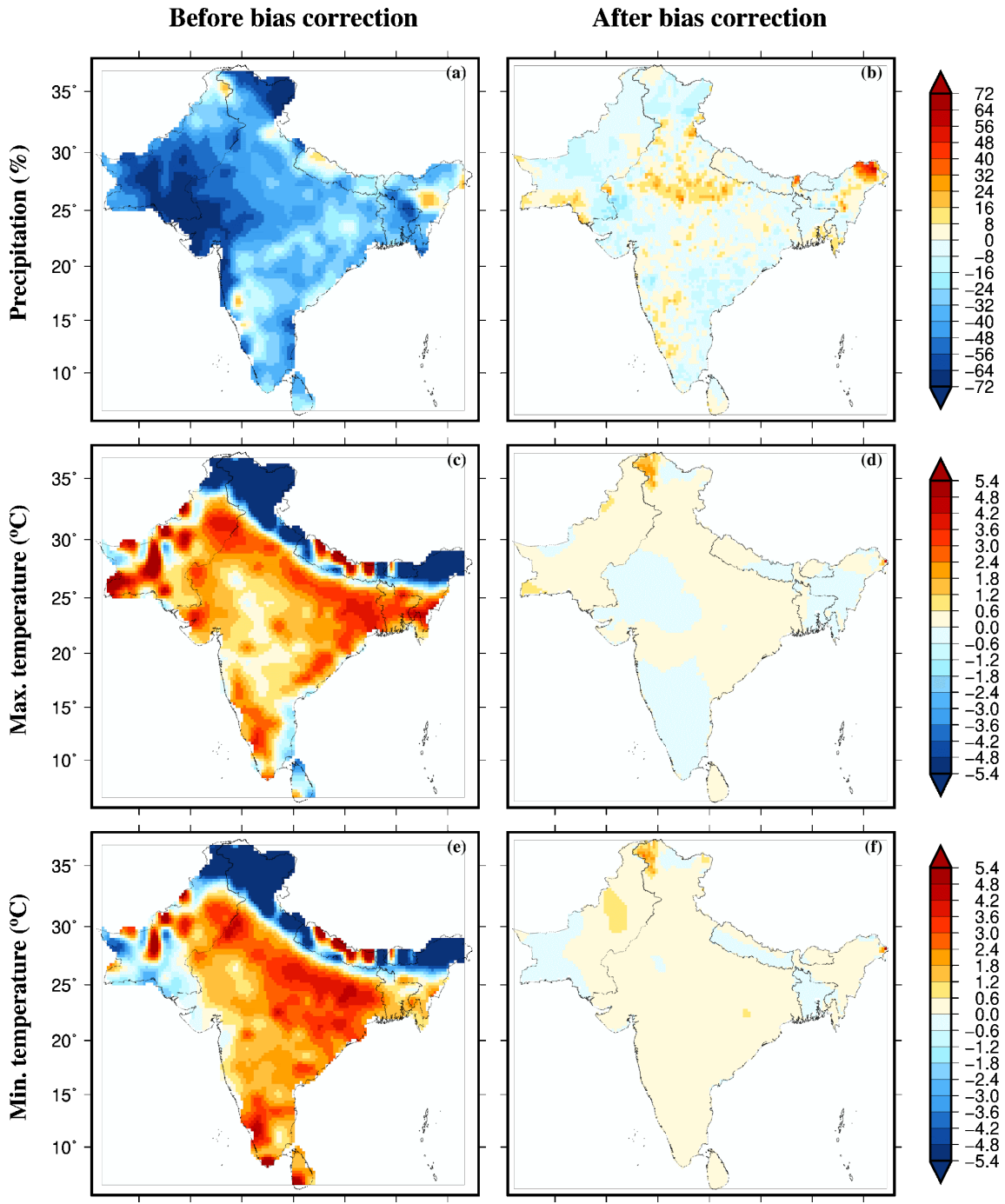

Figure 4. **Multimodel ensemble mean bias in the 90$^{th}$ percentile of precipitation, maximum and minimum temperatures in 13 CMIP6-GCMs.** (a) Bias (%) in extreme precipitation for the historical period (1985-2014), (b) bias in extreme precipitation (%) after the bias correction, (c,d) Bias (°C) in extreme maximum temperature before and after bias correction, and (e,f) bias (°C) in extreme minimum temperature before and after bias correction. The 90$^{th}$ percentile of daily precipitation was estimated using rainy days with precipitation more than 1mm.

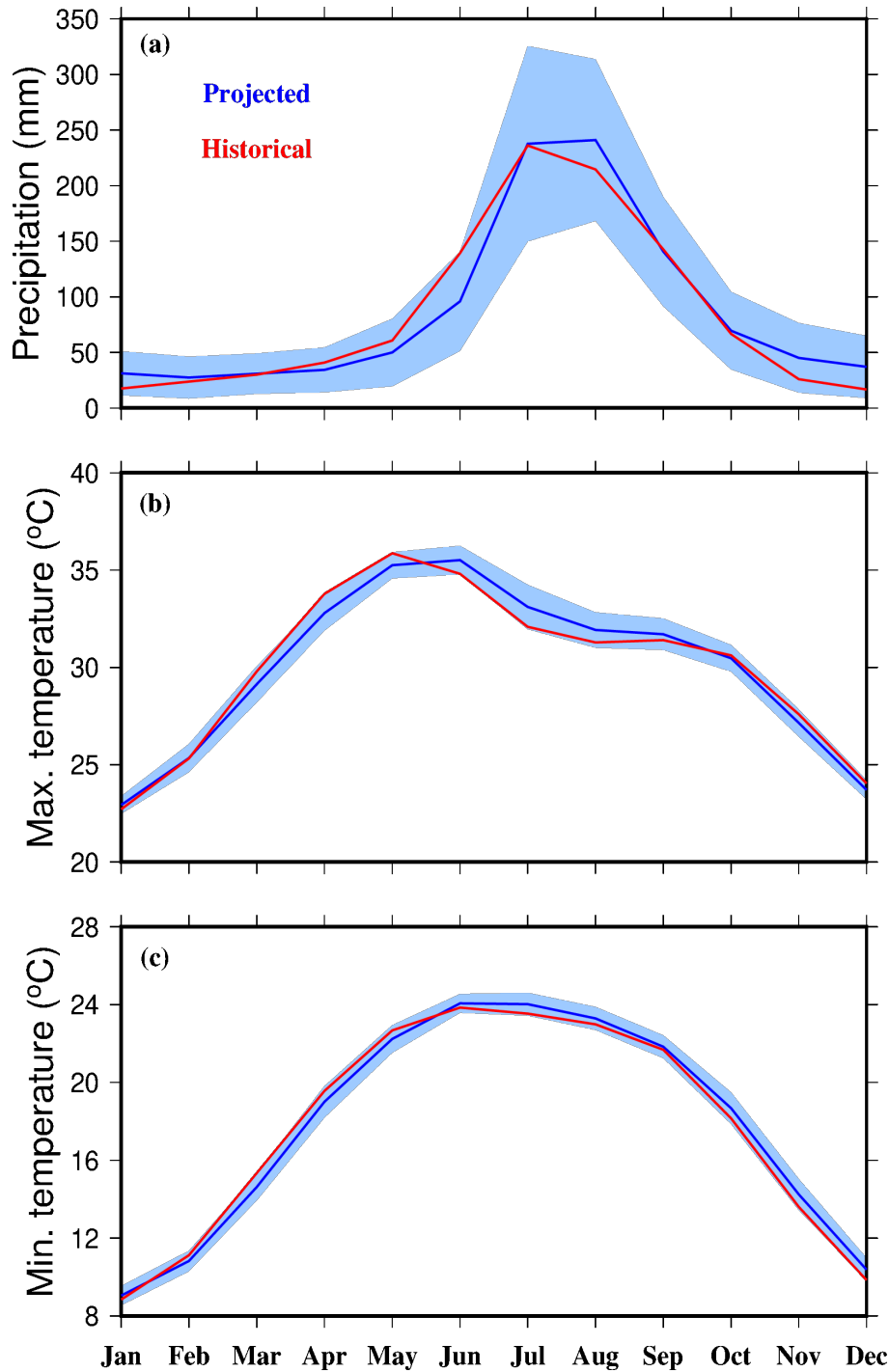

Figure 5. **Seasonal cycle of bias-corrected precipitation, maximum and minimum temperatures.** Comparison of the multimodel model ensemble (blue) mean seasonal cycle of bias-corrected (a) precipitation, (b) maximum temperature, and (c) minimum temperature against the observations for the 1985-2014 period (red). The shaded area represents uncertainty (one standard deviation) of all 13 CMIP6-GCMs.

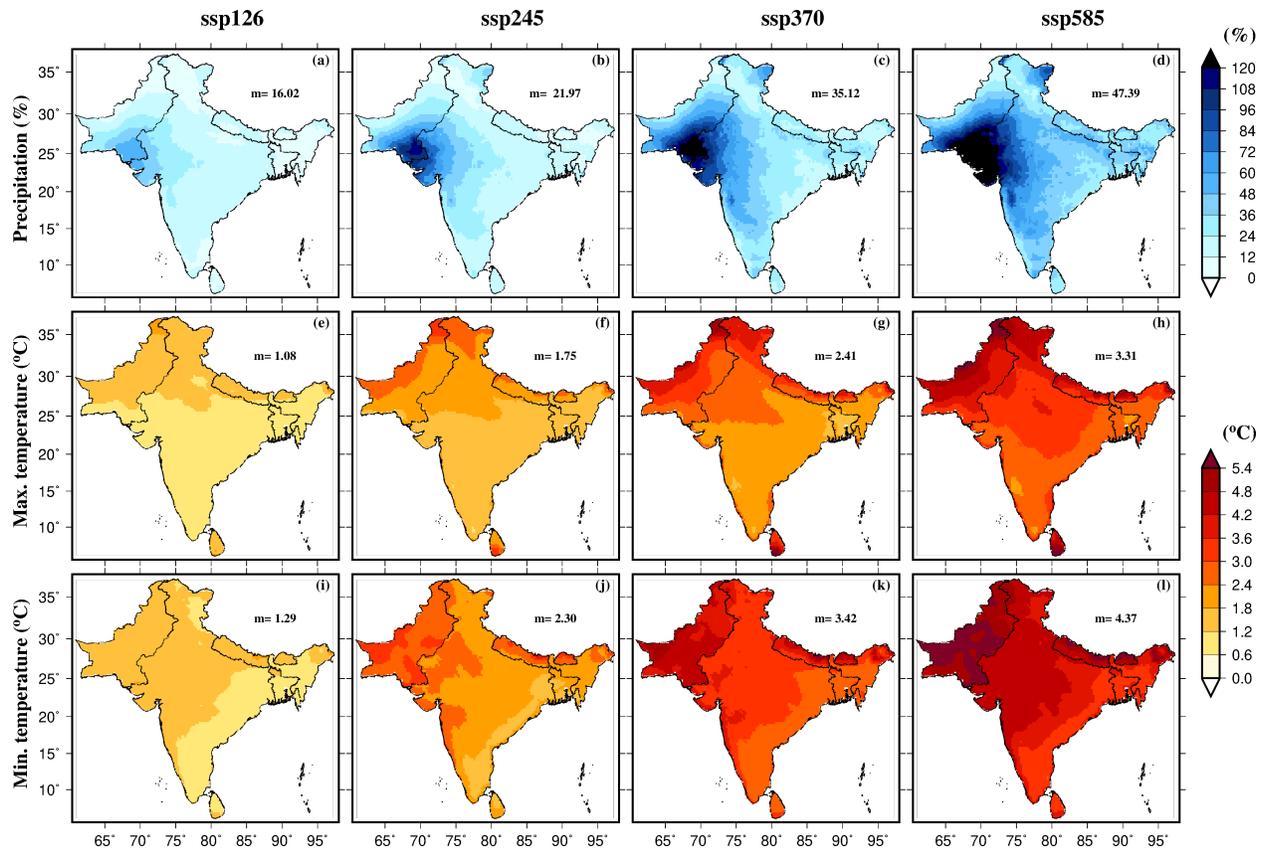

Figure 6. **Projections of precipitation, maximum and minimum temperatures for the end of the 21$^{st}$ century using bias-corrected data from CMIP6-GCMs.** (a-d) The multimodel ensemble mean projected change in mean annual precipitation (%) for the Far (2074-2100) with respect to the historical period (1988-2014), (e-h) same as (a-d) but for the mean annual maximum temperature, (i-l) same as (a-d) but for the mean annual minimum temperature. Median of the multimodel ensemble mean precipitation; maximum and minimum temperatures are shown in each panel. Projected changes were estimated for the four scenarios (SSP126, SSP245, SSP370, and SSP585) against the historical period.

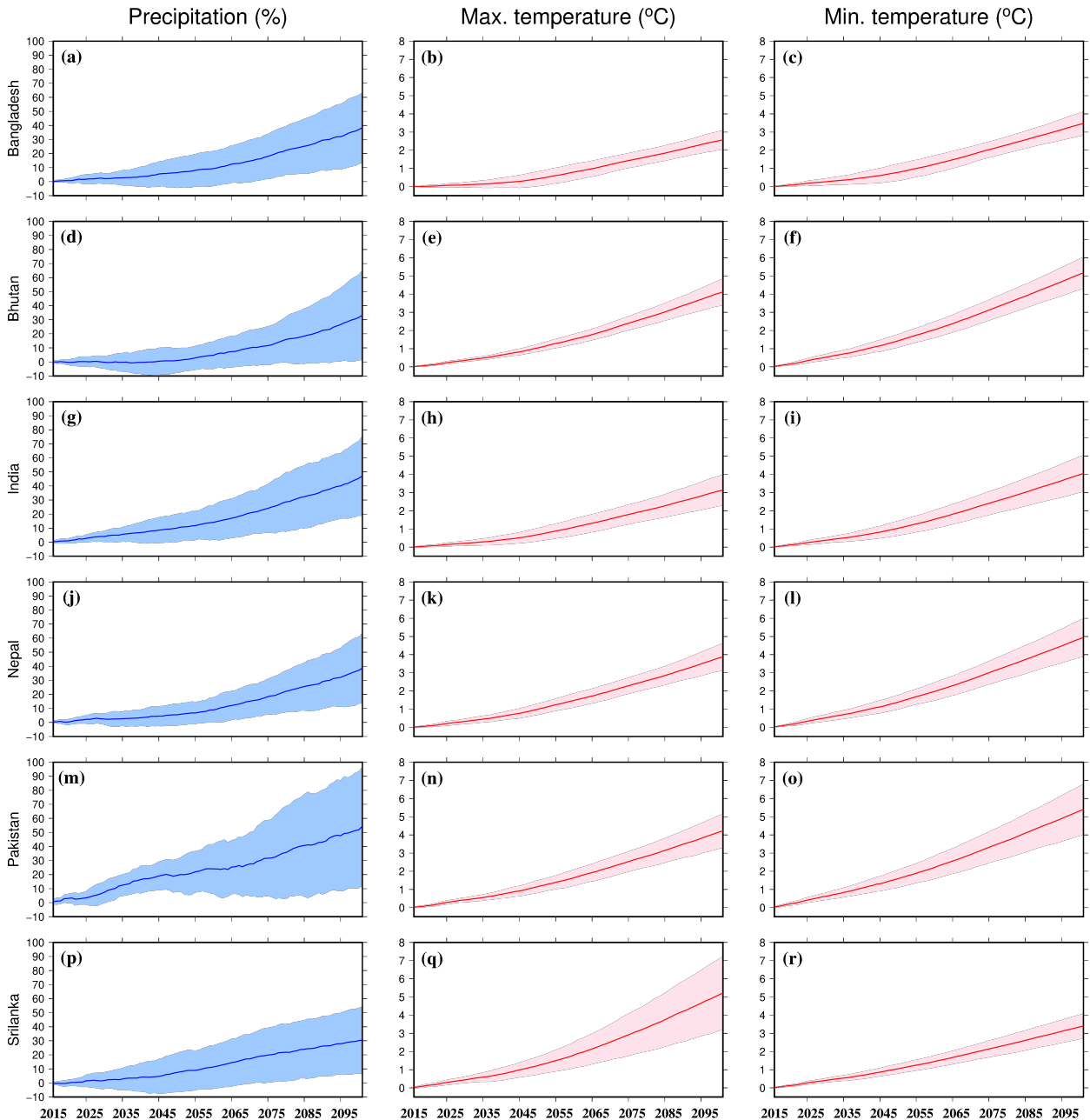

Figure 7. Multimodel **ensemble mean change in precipitation, maximum and minimum temperatures in South Asia.** Countrywise changes in the multimodel ensemble mean annual precipitation (%), maximum temperature (°C), and minimum temperature (°C) estimated using a 30-year moving window against the historical reference period of 1985-2014. The shaded region shows uncertainty (estimated using one standard deviation) based on 13 CMIP6-GCMs.

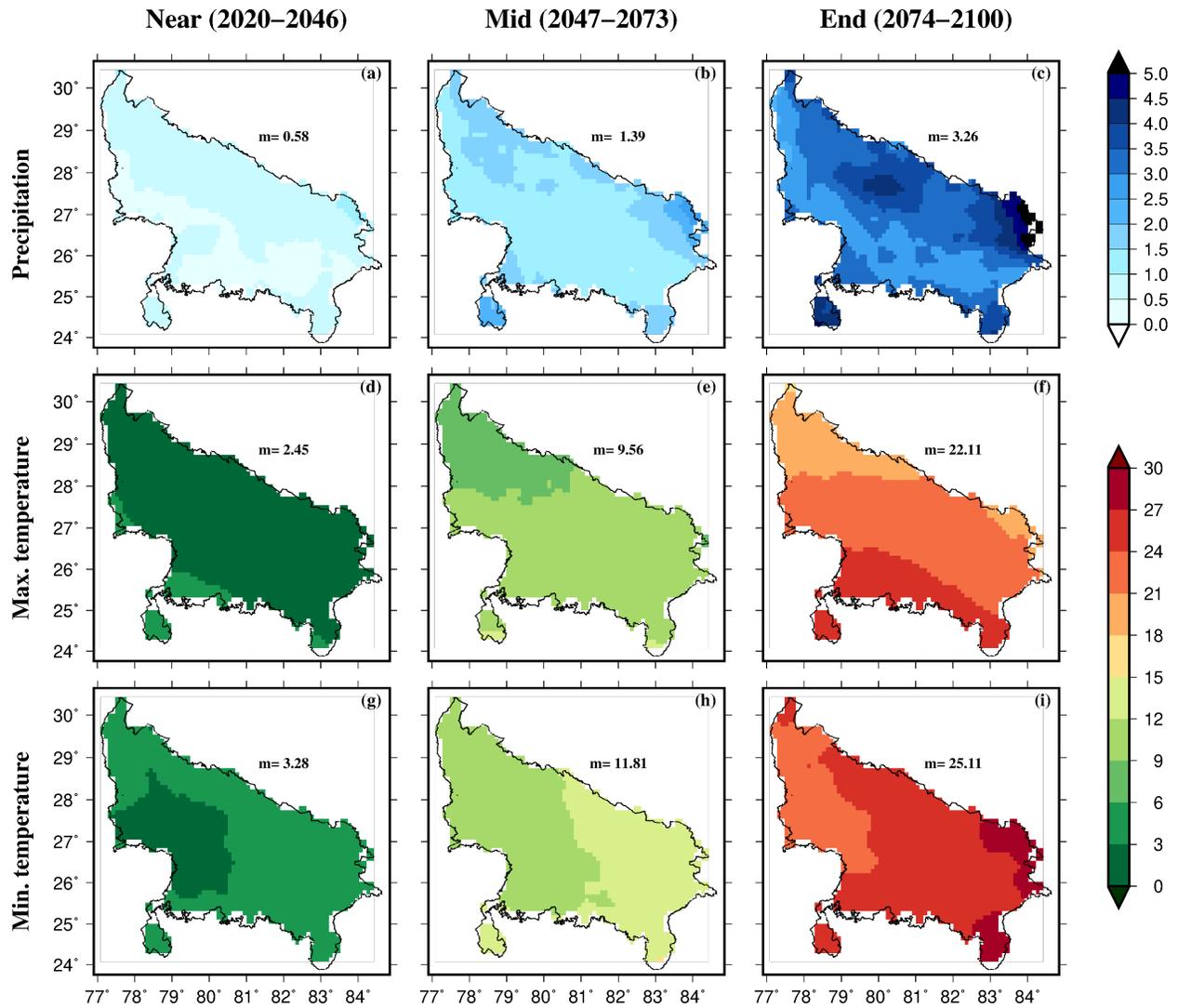

Figure 8. **Changes in the frequency of extreme precipitation, maximum and minimum temperature in the state of Uttar Pradesh.** Projected changes in the frequency of precipitation (a-c), maximum temperature (d-f), and minimum temperature (g-i) extremes estimated using 95the percentile of rainy days (precipitation more than 1mm) and 95$^{th}$ percentile of maximum and minimum summer (April-May) temperatures for the state of Uttar Pradesh (India). Median frequency is shown in each panel. Changes in the frequency were estimated against the historical reference period of 1988-2014.

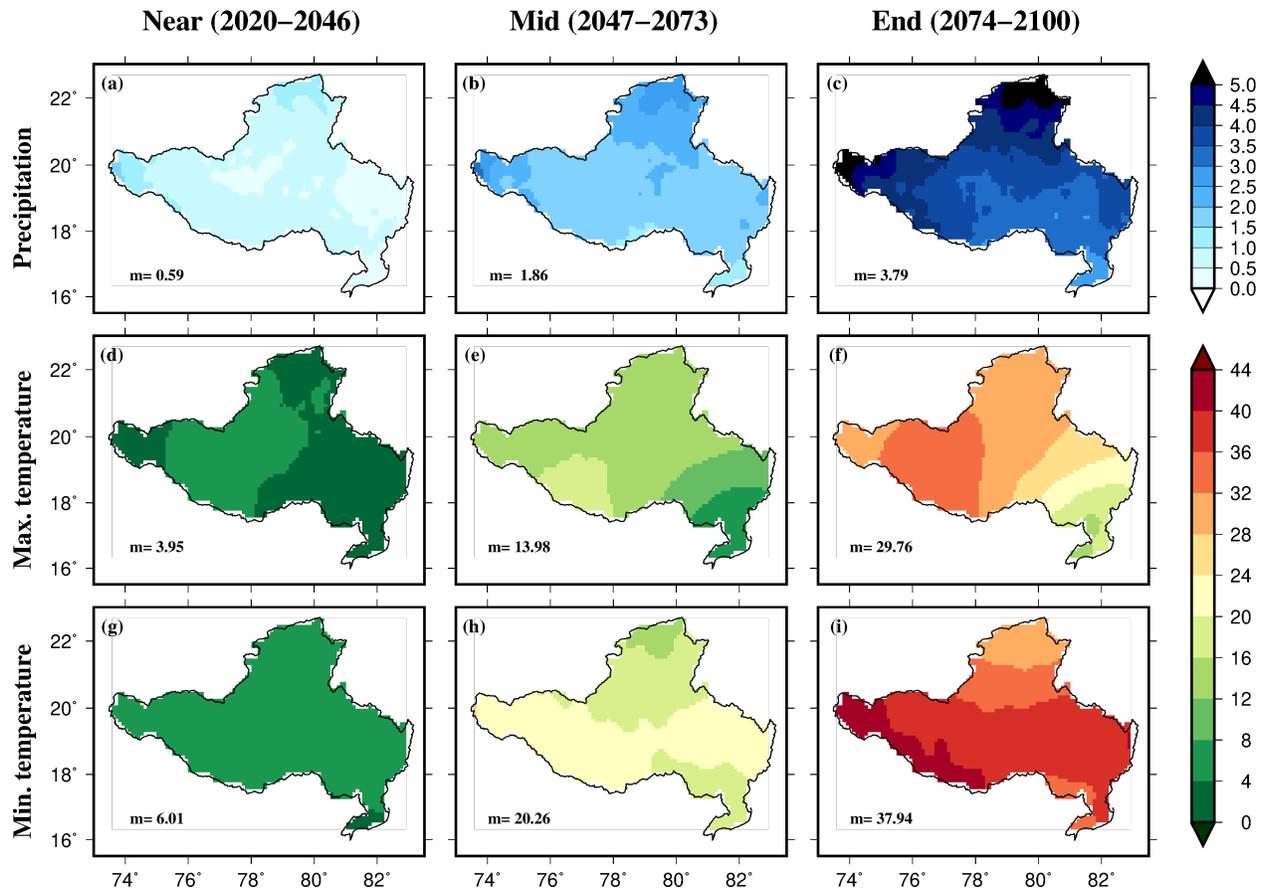

Figure 9. **Changes in the frequency of extreme precipitation, maximum and minimum temperature in the state of Godavari basin.** Projected changes in the frequency of precipitation (a-c), maximum temperature (d-f), and minimum temperature (g-i) extremes estimated using 95[th] percentile of rainy days (precipitation more than 1mm) and 95[th] percentile of maximum and minimum summer (April-May) temperatures for the state of Uttar Pradesh (India). Median frequency is shown in each panel. Changes in the frequency were estimated against the historical reference period of 1988-2014.
References

**Supplemental Information**

**Bias-corrected climate projections from Coupled Model Intercomparison Project-6 (CMIP6) for South Asia**

Vimal Mishra[1,2], Udit Bhatia[1], Amar Deep Tiwari[1]
1--Civil Engineering, Indian Institute of Technology (IIT) Gandhinagar
2—Earth Sciences, Indian Institute of Technology Gandhinagar
*Corresponding Author: vmishra@iitgn.ac.in

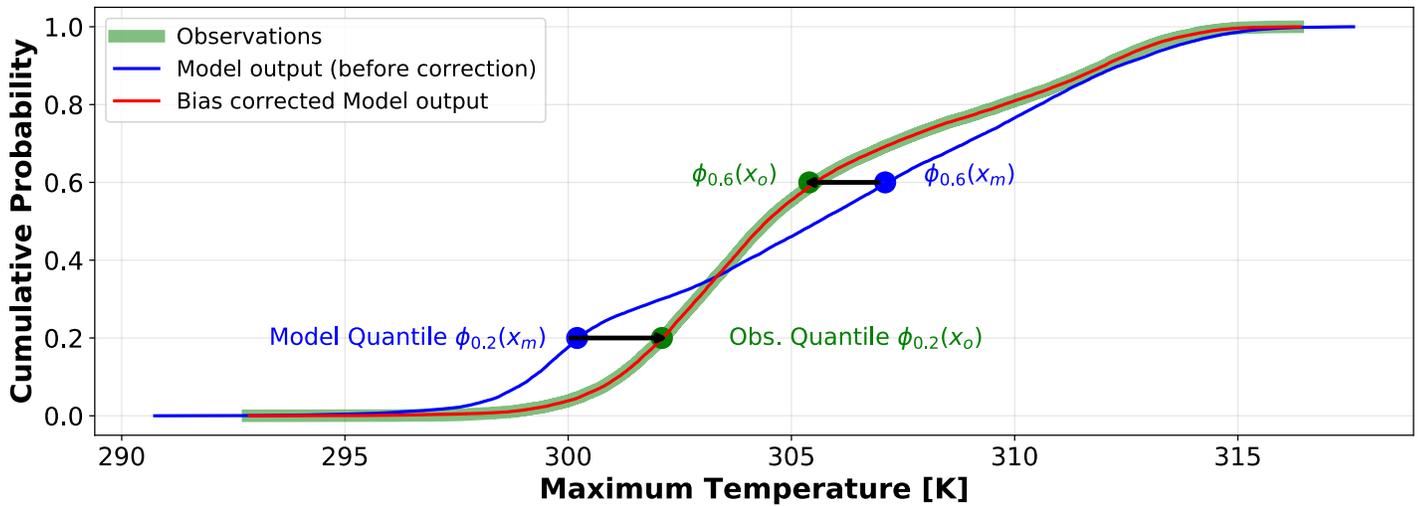

**Figure S1:** Illustration of Quantile-Quantile mapping of model output ($x_m$) (obtained from CMIP6 archive) and observed daily maximum temperature ($x_o$) over a representative grid-point chosen randomly from the Indian Subcontinent. While at 20th percentile ($\phi_{0.2}$), $x_m$ is lower than $x_o$ for the same quantile, $x_m$ exhibits higher bias than $x_o$ at 60th percentile ($\phi_{0.6}$). The empirical adjustment of modeled outputs using non-parametric quantile mapping results in the elimination of systematic bias between $x_m$ and $x_o$.

Table S1. CMIP6 GCMs that are used for bias corrected projections.

| S. No. | Model name | Latitude resolution (degree) | Longitude resolution (degree) |
|---|---|---|---|
| 1 | ACCESS-CM2 | 1.25 | 1.875 |
| 2 | ACCESS-ESM1-5 | 1.25 | 1.875 |
| 3 | BCC-CSM2-MR | 1.1215 | 1.125 |
| 4 | CanESM5 | 2.7906 | 2.8125 |
| 5 | EC-Earth3 | 0.7018 | 0.703125 |
| 6 | EC-Earth3-Veg | 0.7018 | 0.703125 |
| 7 | INM-CM4-8 | 1.5 | 2 |
| 8 | INM-CM5-0 | 1.5 | 2 |
| 9 | MPI-ESM1-2-HR | 0.9351 | 0.9375 |
| 10 | MPI-ESM1-2-LR | 1.8653 | 1.875 |
| 11 | MRI-ESM2-0 | 1.1215 | 1.125 |
| 12 | NorESM2-LM | 1.8947 | 2.5 |
| 13 | NorESM2-MM | 0.9424 | 1.25 |

Table S2: Country wise multimodel ensemble mean projected change in mean annual precipitation (%). Historical mean is provided in mm. Uncertainty (one standard deviation) was estimated using bias corrected data from 13 CMIP6-GCMs.

| Country | | Bangladesh | Bhutan | India | Nepal | Pakistan | Sri Lanka |
|---|---|---|---|---|---|---|---|
| Historic mean (mm) | | 2424.62 | 1921.94 | 1134.79 | 1394.00 | 290.84 | 1741.22 |
| ssp126 | Near | 6.20 ± 9.22 | 3.14 ± 6.65 | 11.40 ± 8.26 | 7.46 ± 7.31 | 19.94 ± 10.28 | 5.32 ± 12.09 |
| | Mid | 10.92 ± 10.52 | 6.65 ± 10.20 | 16.09 ± 8.82 | 11.27 ± 7.40 | 18.55 ± 12.56 | 11.51 ± 12.70 |
| | Far | 9.98 ± 7.19 | 11.43 ± 9.81 | 16.13 ± 9.12 | 13.25 ± 6.48 | 17.16 ± 14.01 | 11.13 ± 12.63 |
| ssp245 | Near | 3.71 ± 9.68 | 0.67 ± 8.79 | 7.39 ± 6.40 | 4.60 ± 6.83 | 13.00 ± 13.32 | 6.48 ± 9.08 |
| | Mid | 10.92 ± 13.69 | 6.88 ± 12.42 | 16.24 ± 10.23 | 12.02 ± 10.46 | 17.53 ± 15.01 | 14.19 ± 14.12 |
| | Far | 15.05 ± 15.38 | 11.53 ± 13.93 | 22.80 ± 12.23 | 15.16 ± 11.98 | 27.74 ± 17.65 | 18.84 ± 16.23 |
| ssp370 | Near | 3.66 ± 7.28 | 2.19 ± 9.56 | 9.76 ± 7.63 | 6.63 ± 8.19 | 17.68 ± 15.66 | 7.82 ± 11.20 |
| | Mid | 11.68 ± 13.48 | 7.76 ± 15.19 | 17.94 ± 15.69 | 13.59 ± 15.55 | 24.43 ± 18.01 | 12.73 ± 18.77 |
| | Far | 27.02 ± 20.00 | 20.85 ± 22.63 | 34.53 ± 20.35 | 26.42 ± 18.28 | 45.86 ± 28.64 | 23.83 ± 20.89 |
| ssp585 | Near | 6.19 ± 10.54 | 0.91 ± 9.71 | 9.78 ± 10.21 | 5.48 ± 7.41 | 20.44 ± 11.87 | 6.50 ± 14.01 |
| | Mid | 18.11 ± 16.05 | 12.08 ± 13.79 | 24.30 ± 17.81 | 18.38 ± 12.78 | 31.98 ± 27.75 | 20.02 ± 19.87 |
| | Far | 39.46 ± 25.37 | 35.29 ± 33.75 | 48.67 ± 28.66 | 39.96 ± 26.13 | 53.58 ± 42.50 | 31.79 ± 23.12 |

Table S3: Country wise multimodel projected change in mean annual maximum temperature (°C). Uncertainty (one standard deviation) was estimated using bias corrected data from 13 CMIP6-GCMs.

| Country | | Bangladesh | Bhutan | India | Nepal | Pakistan | Sri Lanka |
|---|---|---|---|---|---|---|---|
| Historic mean | | 30.35 | 17.01 | 30.67 | 19.93 | 29.44 | 30.61 |
| ssp126 | Near | 0.48 ± 0.18 | 0.84 ± 0.17 | 0.58 ± 0.24 | 0.82 ± 0.22 | 0.90 ± 0.26 | 0.97 ± 0.37 |
| | Mid | 0.90 ± 0.17 | 1.30 ± 0.31 | 1.02 ± 0.38 | 1.28 ± 0.35 | 1.35 ± 0.42 | 1.46 ± 0.60 |
| | Far | 0.95 ± 0.24 | 1.31 ± 0.40 | 1.05 ± 0.40 | 1.32 ± 0.45 | 1.39 ± 0.46 | 1.49 ± 0.64 |
| ssp245 | Near | 0.33 ± 0.27 | 0.82 ± 0.16 | 0.56 ± 0.27 | 0.81 ± 0.21 | 0.94 ± 0.25 | 0.97 ± 0.44 |
| | Mid | 1.00 ± 0.26 | 1.66 ± 0.31 | 1.28 ± 0.44 | 1.64 ± 0.34 | 1.80 ± 0.45 | 1.97 ± 0.74 |
| | Far | 1.42 ± 0.32 | 2.20 ± 0.45 | 1.72 ± 0.54 | 2.18 ± 0.45 | 2.36 ± 0.58 | 2.59 ± 0.96 |
| ssp370 | Near | 0.29 ± 0.31 | 0.81 ± 0.23 | 0.48 ± 0.36 | 0.77 ± 0.28 | 0.91 ± 0.34 | 1.03 ± 0.45 |
| | Mid | 0.95 ± 0.40 | 1.86 ± 0.30 | 1.37 ± 0.49 | 1.81 ± 0.36 | 2.05 ± 0.54 | 2.45 ± 0.93 |
| | Far | 1.85 ± 0.53 | 3.18 ± 0.58 | 2.40 ± 0.70 | 3.02 ± 0.57 | 3.31 ± 0.74 | 4.20 ± 1.59 |
| ssp585 | Near | 0.36 ± 0.39 | 0.97 ± 0.20 | 0.61 ± 0.35 | 0.91 ± 0.29 | 1.07 ± 0.33 | 1.13 ± 0.46 |
| | Mid | 1.45 ± 0.41 | 2.42 ± 0.39 | 1.81 ± 0.54 | 2.31 ± 0.45 | 2.55 ± 0.55 | 2.95 ± 1.16 |
| | Far | 2.64 ± 0.55 | 4.22 ± 0.75 | 3.22 ± 0.88 | 3.97 ± 0.77 | 4.33 ± 0.98 | 5.34 ± 2.05 |

Table S4: Same as Table S3 but for mean annual minimum temperature. Uncertainty (one standard deviation) was estimated using bias corrected data from 13 CMIP6-GCMs.

| Country | | Bangladesh | Bhutan | India | Nepal | Pakistan | Sri Lanka |
|---|---|---|---|---|---|---|---|
| **Historic mean** | | 21.00 | 5.66 | 18.68 | 7.57 | 15.52 | 23.31 |
| **ssp126** | **Near** | 0.69 ± 0.20 | 1.15 ± 0.28 | 0.84 ± 0.29 | 1.12 ± 0.29 | 1.27 ± 0.39 | 0.82 ± 0.23 |
| | **Mid** | 1.11 ± 0.27 | 1.61 ± 0.48 | 1.25 ± 0.45 | 1.55 ± 0.45 | 1.67 ± 0.53 | 1.15 ± 0.35 |
| | **Far** | 1.14 ± 0.29 | 1.58 ± 0.55 | 1.25 ± 0.48 | 1.54 ± 0.51 | 1.65 ± 0.59 | 1.13 ± 0.37 |
| **ssp245** | **Near** | 0.65 ± 0.33 | 1.20 ± 0.27 | 0.84 ± 0.33 | 1.14 ± 0.30 | 1.28 ± 0.38 | 0.85 ± 0.24 |
| | **Mid** | 1.41 ± 0.37 | 2.24 ± 0.49 | 1.68 ± 0.50 | 2.10 ± 0.51 | 2.32 ± 0.60 | 1.54 ± 0.37 |
| | **Far** | 1.88 ± 0.42 | 2.93 ± 0.52 | 2.19 ± 0.63 | 2.73 ± 0.57 | 2.97 ± 0.80 | 1.91 ± 0.45 |
| **ssp370** | **Near** | 0.68 ± 0.36 | 1.28 ± 0.36 | 0.82 ± 0.36 | 1.17 ± 0.39 | 1.30 ± 0.44 | 0.89 ± 0.25 |
| | **Mid** | 1.64 ± 0.51 | 2.75 ± 0.57 | 1.96 ± 0.62 | 2.53 ± 0.66 | 2.72 ± 0.77 | 1.81 ± 0.43 |
| | **Far** | 2.78 ± 0.64 | 4.41 ± 0.73 | 3.25 ± 0.87 | 4.06 ± 0.90 | 4.38 ± 1.14 | 2.82 ± 0.61 |
| **ssp585** | **Near** | 0.71 ± 0.46 | 1.34 ± 0.35 | 0.96 ± 0.39 | 1.29 ± 0.38 | 1.50 ± 0.47 | 0.99 ± 0.26 |
| | **Mid** | 2.06 ± 0.43 | 3.15 ± 0.57 | 2.43 ± 0.65 | 3.03 ± 0.64 | 3.35 ± 0.86 | 2.16 ± 0.46 |
| | **Far** | 3.56 ± 0.68 | 5.30 ± 0.90 | 4.14 ± 1.03 | 5.06 ± 1.08 | 5.52 ± 1.44 | 3.47 ± 0.70 |

**Table S5:** Basin wise multimodel ensemble mean projected change in mean annual precipitation (%) Uncertainty (one standard deviation) was estimated using bias corrected data from 13 CMIP6-GCMs.

| Basin | Historic mean (mm) | ssp126 | | | ssp245 | | | ssp370 | | | ssp585 | | |
|---|---|---|---|---|---|---|---|---|---|---|---|---|---|
| | | Near | Mid | Far | Near | Mid | Far | Near | Mid | Far | Near | Mid | Far |
| Bhahmani | 1452.06 | 8.11 ± 9.77 | 14.44 ± 9.79 | 16.95 ± 9.78 | 2.24 ± 5.95 | 9.84 ± 5.73 | 16.60 ± 11.26 | 3.16 ± 8.16 | 8.10 ± 13.52 | 20.57 ± 14.45 | 3.95 ± 9.02 | 16.95 ± 12.99 | 33.13 ± 16.99 |
| Brahmaputra | 1692.12 | 4.18 ± 5.06 | 8.47 ± 7.65 | 11.15 ± 7.50 | 2.15 ± 7.23 | 9.44 ± 8.74 | 13.47 ± 9.33 | 3.56 ± 6.60 | 10.90 ± 9.88 | 25.91 ± 15.16 | 3.22 ± 6.15 | 15.89 ± 9.56 | 39.04 ± 22.11 |
| Cauvery | 942.30 | 8.33 ± 11.95 | 14.53 ± 10.93 | 12.95 ± 10.89 | 7.42 ± 8.38 | 15.71 ± 10.76 | 19.08 ± 12.20 | 8.12 ± 7.59 | 15.19 ± 14.04 | 31.71 ± 17.25 | 9.57 ± 12.05 | 24.24 ± 18.76 | 43.05 ± 21.97 |
| East Coast | 1015.93 | 8.19 ± 13.02 | 13.27 ± 8.84 | 10.52 ± 9.53 | 7.59 ± 8.82 | 13.07 ± 8.59 | 20.18 ± 14.01 | 7.57 ± 10.33 | 15.95 ± 12.45 | 30.76 ± 20.96 | 10.20 ± 11.76 | 23.33 ± 16.42 | 42.59 ± 29.57 |
| Ganga | 1147.34 | 10.08 ± 10.36 | 15.57 ± 9.25 | 15.73 ± 8.92 | 6.95 ± 7.78 | 15.05 ± 11.38 | 21.55 ± 15.19 | 8.73 ± 8.10 | 17.57 ± 16.09 | 32.37 ± 19.43 | 9.04 ± 9.75 | 22.65 ± 17.02 | 46.40 ± 27.44 |
| Godavari | 1152.89 | 14.11 ± 10.04 | 16.84 ± 11.70 | 17.34 ± 13.51 | 8.31 ± 7.19 | 17.38 ± 18.31 | 22.61 ± 16.37 | 10.60 ± 14.69 | 18.08 ± 25.07 | 34.41 ± 33.02 | 9.11 ± 16.89 | 24.14 ± 27.02 | 47.68 ± 45.24 |
| Indus | 525.11 | 13.26 ± 7.76 | 13.97 ± 8.22 | 12.56 ± 9.15 | 8.34 ± 9.03 | 13.45 ± 9.84 | 19.49 ± 11.55 | 11.66 ± 8.74 | 17.47 ± 12.42 | 33.94 ± 17.81 | 13.38 ± 8.87 | 23.32 ± 17.47 | 42.63 ± 23.93 |
| Krishna | 822.30 | 16.94 ± 14.17 | 19.15 ± 14.18 | 18.51 ± 19.63 | 11.33 ± 13.15 | 21.34 ± 23.29 | 26.50 ± 25.66 | 14.33 ± 16.98 | 23.61 ± 29.99 | 42.72 ± 43.57 | 14.38 ± 20.14 | 30.05 ± 37.47 | 53.98 ± 57.17 |
| Mahanadi | 1341.97 | 11.70 ± 10.18 | 16.46 ± 9.59 | 18.92 ± 10.16 | 5.49 ± 5.43 | 12.52 ± 10.70 | 19.73 ± 11.22 | 6.32 ± 9.77 | 12.14 ± 18.64 | 25.19 ± 20.50 | 5.46 ± 11.08 | 19.45 ± 17.62 | 38.62 ± 27.25 |
| Mahi | 880.62 | 25.68 ± 18.23 | 36.65 ± 23.40 | 32.59 ± 23.37 | 17.08 ± 19.71 | 37.21 ± 24.76 | 52.78 ± 30.10 | 25.43 ± 20.88 | 40.27 ± 39.92 | 67.94 ± 52.62 | 25.53 ± 22.98 | 53.21 ± 45.83 | 101.03 ± 84.23 |
| Narmada | 1126.08 | 18.94 ± 12.21 | 24.84 ± 14.13 | 23.70 ± 16.03 | 13.45 ± 12.69 | 26.38 ± 21.69 | 34.01 ± 19.75 | 16.38 ± 15.86 | 27.61 ± 31.84 | 45.64 ± 40.55 | 15.21 ± 18.76 | 35.76 ± 35.40 | 66.89 ± 59.08 |
| North East Coast | 1243.27 | 8.22 ± 8.22 | 11.93 ± 8.32 | 14.55 ± 7.47 | 1.94 ± 5.73 | 9.43 ± 8.36 | 13.50 ± 7.28 | 2.67 ± 6.68 | 5.99 ± 11.66 | 17.62 ± 14.22 | 2.38 ± 10.30 | 13.84 ± 12.43 | 29.46 ± 20.36 |
| Pennar | 777.74 | 12.81 ± 14.25 | 16.16 ± 9.20 | 14.26 ± 13.71 | 7.95 ± 11.26 | 14.82 ± 16.61 | 24.04 ± 21.49 | 8.89 ± 14.91 | 19.13 ± 20.90 | 37.09 ± 35.00 | 11.21 ± 16.27 | 25.83 ± 27.07 | 49.98 ± 40.28 |
| Sabarmati | 512.99 | 34.56 ± 25.48 | 43.63 ± 29.93 | 40.43 ± 25.48 | 24.41 ± 26.73 | 45.93 ± 27.36 | 73.83 ± 38.16 | 34.41 ± 24.35 | 51.81 ± 43.08 | 91.08 ± 56.88 | 36.31 ± 30.66 | 67.65 ± 50.14 | 130.00 ± 101.03 |
| South Coast | 1395.07 | 4.71 ± 12.14 | 12.36 ± 11.83 | 11.41 ± 11.70 | 5.45 ± 8.59 | 13.67 ± 11.96 | 17.76 ± 13.96 | 7.78 ± 9.42 | 12.85 ± 17.56 | 28.38 ± 21.43 | 6.83 ± 12.84 | 21.18 ± 19.97 | 38.46 ± 26.86 |
| Subarnarekha | 1428.76 | 7.27 ± 9.19 | 13.65 ± 9.85 | 14.28 ± 8.89 | 2.13 ± 7.57 | 9.55 ± 6.61 | 16.26 ± 13.28 | 2.84 ± 6.05 | 9.05 ± 11.08 | 21.69 ± 14.51 | 4.74 ± 8.02 | 16.54 ± 11.53 | 34.37 ± 17.33 |
| Tapi | 954.05 | 21.40 ± 12.85 | 26.07 ± 18.19 | 23.70 ± 18.32 | 15.31 ± 13.40 | 30.08 ± 25.86 | 37.87 ± 22.98 | 20.09 ± 19.91 | 32.17 ± 36.79 | 54.67 ± 49.96 | 18.26 ± 22.46 | 39.85 ± 41.76 | 76.12 ± 73.28 |
| West Coast | 2718.34 | 13.32 ± 11.58 | 16.27 ± 13.36 | 15.77 ± 15.55 | 9.49 ± 11.25 | 19.51 ± 17.69 | 22.88 ± 20.04 | 12.92 ± 12.99 | 18.82 ± 22.85 | 32.60 ± 31.04 | 12.94 ± 16.95 | 25.48 ± 29.57 | 42.83 ± 43.31 |

Table S6: Same as Table S5 but for mean annual maximum temperature (°C)

| Basin | Historic mean | ssp126 | | | ssp245 | | | ssp370 | | | ssp585 | | |
|---|---|---|---|---|---|---|---|---|---|---|---|---|---|
| | | Near | Mid | Far | Near | Mid | Far | Near | Mid | Far | Near | Mid | Far |
| Bhahmani | 32.08 | 0.43 ± 0.20 | 0.86 ± 0.28 | 0.90 ± 0.29 | 0.35 ± 0.30 | 1.05 ± 0.35 | 1.43 ± 0.43 | 0.30 ± 0.36 | 1.07 ± 0.46 | 2.04 ± 0.66 | 0.35 ± 0.40 | 1.52 ± 0.48 | 2.88 ± 0.78 |
| Brahmaputra | 18.41 | 0.81 ± 0.17 | 1.23 ± 0.27 | 1.24 ± 0.36 | 0.81 ± 0.15 | 1.56 ± 0.30 | 2.05 ± 0.40 | 0.80 ± 0.24 | 1.78 ± 0.33 | 2.93 ± 0.51 | 0.95 ± 0.23 | 2.25 ± 0.35 | 3.79 ± 0.59 |
| Cauvery | 29.78 | 0.51 ± 0.27 | 0.80 ± 0.34 | 0.84 ± 0.36 | 0.48 ± 0.33 | 1.05 ± 0.41 | 1.36 ± 0.54 | 0.47 ± 0.35 | 1.20 ± 0.49 | 2.01 ± 0.73 | 0.53 ± 0.34 | 1.50 ± 0.54 | 2.57 ± 0.92 |
| East Coast | 32.75 | 0.56 ± 0.32 | 0.93 ± 0.37 | 0.99 ± 0.36 | 0.52 ± 0.39 | 1.23 ± 0.46 | 1.57 ± 0.59 | 0.49 ± 0.48 | 1.36 ± 0.62 | 2.32 ± 0.86 | 0.55 ± 0.46 | 1.72 ± 0.66 | 3.01 ± 1.03 |
| Ganga | 28.39 | 0.62 ± 0.28 | 1.11 ± 0.39 | 1.17 ± 0.42 | 0.55 ± 0.27 | 1.37 ± 0.42 | 1.85 ± 0.51 | 0.45 ± 0.40 | 1.38 ± 0.47 | 2.49 ± 0.67 | 0.60 ± 0.40 | 1.92 ± 0.55 | 3.47 ± 0.84 |
| Godavari | 32.89 | 0.42 ± 0.28 | 0.88 ± 0.44 | 0.88 ± 0.43 | 0.45 ± 0.35 | 1.09 ± 0.56 | 1.49 ± 0.63 | 0.33 ± 0.44 | 1.16 ± 0.61 | 2.14 ± 0.88 | 0.47 ± 0.43 | 1.57 ± 0.66 | 2.91 ± 1.09 |
| Indus | 25.32 | 0.93 ± 0.29 | 1.39 ± 0.45 | 1.43 ± 0.49 | 0.98 ± 0.27 | 1.84 ± 0.48 | 2.41 ± 0.62 | 0.94 ± 0.34 | 2.10 ± 0.59 | 3.38 ± 0.81 | 1.10 ± 0.35 | 2.60 ± 0.61 | 4.40 ± 1.04 |
| Krishna | 32.11 | 0.43 ± 0.29 | 0.82 ± 0.41 | 0.83 ± 0.42 | 0.44 ± 0.34 | 1.03 ± 0.53 | 1.38 ± 0.62 | 0.37 ± 0.39 | 1.14 ± 0.57 | 2.00 ± 0.84 | 0.46 ± 0.39 | 1.47 ± 0.63 | 2.65 ± 1.05 |
| Mahanadi | 32.59 | 0.43 ± 0.20 | 0.88 ± 0.36 | 0.90 ± 0.34 | 0.41 ± 0.32 | 1.12 ± 0.41 | 1.50 ± 0.49 | 0.31 ± 0.41 | 1.15 ± 0.51 | 2.17 ± 0.72 | 0.42 ± 0.42 | 1.59 ± 0.54 | 3.00 ± 0.88 |
| Mahi | 33.21 | 0.53 ± 0.41 | 0.94 ± 0.58 | 0.99 ± 0.60 | 0.56 ± 0.43 | 1.20 ± 0.70 | 1.60 ± 0.86 | 0.39 ± 0.52 | 1.29 ± 0.75 | 2.29 ± 1.04 | 0.58 ± 0.51 | 1.70 ± 0.83 | 3.03 ± 1.29 |
| Narmada | 32.82 | 0.48 ± 0.37 | 0.94 ± 0.53 | 0.99 ± 0.52 | 0.49 ± 0.39 | 1.19 ± 0.65 | 1.61 ± 0.74 | 0.33 ± 0.52 | 1.22 ± 0.71 | 2.28 ± 0.97 | 0.53 ± 0.49 | 1.69 ± 0.74 | 3.11 ± 1.16 |
| North East Coast | 31.61 | 0.38 ± 0.17 | 0.78 ± 0.28 | 0.77 ± 0.27 | 0.39 ± 0.27 | 0.99 ± 0.32 | 1.32 ± 0.38 | 0.33 ± 0.30 | 1.07 ± 0.40 | 1.96 ± 0.56 | 0.40 ± 0.30 | 1.43 ± 0.43 | 2.65 ± 0.73 |
| Pennar | 32.73 | 0.47 ± 0.30 | 0.86 ± 0.39 | 0.89 ± 0.40 | 0.46 ± 0.36 | 1.12 ± 0.49 | 1.45 ± 0.61 | 0.40 ± 0.44 | 1.22 ± 0.60 | 2.13 ± 0.86 | 0.47 ± 0.41 | 1.57 ± 0.65 | 2.84 ± 1.07 |
| Sabarmati | 33.09 | 0.62 ± 0.31 | 1.01 ± 0.47 | 1.05 ± 0.51 | 0.65 ± 0.32 | 1.33 ± 0.53 | 1.73 ± 0.67 | 0.55 ± 0.37 | 1.45 ± 0.57 | 2.46 ± 0.79 | 0.70 ± 0.40 | 1.84 ± 0.63 | 3.17 ± 0.99 |
| South Coast | 31.53 | 0.62 ± 0.25 | 0.90 ± 0.33 | 0.95 ± 0.36 | 0.59 ± 0.30 | 1.20 ± 0.40 | 1.56 ± 0.52 | 0.59 ± 0.30 | 1.43 ± 0.46 | 2.32 ± 0.68 | 0.67 ± 0.32 | 1.73 ± 0.55 | 2.90 ± 0.88 |
| Subarnarekha | 31.66 | 0.44 ± 0.22 | 0.90 ± 0.27 | 0.97 ± 0.30 | 0.27 ± 0.32 | 1.03 ± 0.34 | 1.45 ± 0.45 | 0.23 ± 0.37 | 0.96 ± 0.47 | 1.96 ± 0.68 | 0.29 ± 0.46 | 1.51 ± 0.52 | 2.87 ± 0.79 |
| Tapi | 33.91 | 0.46 ± 0.38 | 0.92 ± 0.51 | 0.95 ± 0.52 | 0.49 ± 0.40 | 1.12 ± 0.69 | 1.54 ± 0.77 | 0.33 ± 0.49 | 1.19 ± 0.71 | 2.17 ± 0.99 | 0.53 ± 0.46 | 1.60 ± 0.75 | 2.93 ± 1.19 |
| West Coast | 31.43 | 0.59 ± 0.29 | 0.96 ± 0.44 | 0.98 ± 0.46 | 0.59 ± 0.30 | 1.22 ± 0.53 | 1.63 ± 0.65 | 0.55 ± 0.31 | 1.40 ± 0.52 | 2.37 ± 0.82 | 0.66 ± 0.34 | 1.75 ± 0.61 | 3.02 ± 1.04 |

Table S7: Same as Table S5 but for mean annual minimum temperature (°C)

| Basin | Historic mean | ssp126 | | | ssp245 | | | ssp370 | | | ssp585 | | |
|---|---|---|---|---|---|---|---|---|---|---|---|---|---|
| | | Near | Mid | Far | Near | Mid | Far | Near | Mid | Far | Near | Mid | Far |
| Bhahmani | 20.85 | 0.64 ± 0.17 | 1.06 ± 0.33 | 1.09 ± 0.34 | 0.61 ± 0.30 | 1.40 ± 0.39 | 1.85 ± 0.43 | 0.60 ± 0.34 | 1.63 ± 0.47 | 2.83 ± 0.61 | 0.67 ± 0.38 | 2.07 ± 0.46 | 3.71 ± 0.67 |
| Brahmaputra | 6.42 | 1.04 ± 0.25 | 1.47 ± 0.40 | 1.44 ± 0.46 | 1.07 ± 0.25 | 1.97 ± 0.44 | 2.55 ± 0.51 | 1.13 ± 0.36 | 2.38 ± 0.55 | 3.81 ± 0.74 | 1.20 ± 0.35 | 2.77 ± 0.54 | 4.67 ± 0.94 |
| Cauvery | 19.63 | 0.68 ± 0.26 | 0.98 ± 0.42 | 0.99 ± 0.41 | 0.70 ± 0.27 | 1.38 ± 0.46 | 1.75 ± 0.59 | 0.74 ± 0.28 | 1.66 ± 0.49 | 2.64 ± 0.79 | 0.81 ± 0.29 | 1.99 ± 0.57 | 3.25 ± 1.00 |
| East Coast | 22.61 | 0.67 ± 0.27 | 1.01 ± 0.45 | 1.02 ± 0.44 | 0.68 ± 0.29 | 1.39 ± 0.50 | 1.76 ± 0.61 | 0.72 ± 0.33 | 1.67 ± 0.53 | 2.66 ± 0.80 | 0.78 ± 0.35 | 2.01 ± 0.59 | 3.27 ± 1.01 |
| Ganga | 15.82 | 0.90 ± 0.31 | 1.32 ± 0.46 | 1.34 ± 0.50 | 0.86 ± 0.36 | 1.76 ± 0.53 | 2.32 ± 0.65 | 0.84 ± 0.41 | 2.04 ± 0.69 | 3.42 ± 0.96 | 0.99 ± 0.44 | 2.56 ± 0.70 | 4.42 ± 1.12 |
| Godavari | 20.75 | 0.74 ± 0.33 | 1.21 ± 0.50 | 1.21 ± 0.50 | 0.75 ± 0.38 | 1.62 ± 0.57 | 2.11 ± 0.68 | 0.71 ± 0.40 | 1.86 ± 0.65 | 3.15 ± 0.90 | 0.87 ± 0.42 | 2.35 ± 0.70 | 4.05 ± 1.08 |
| Indus | 11.95 | 1.16 ± 0.36 | 1.55 ± 0.50 | 1.52 ± 0.55 | 1.18 ± 0.36 | 2.12 ± 0.58 | 2.73 ± 0.75 | 1.19 ± 0.43 | 2.51 ± 0.77 | 4.04 ± 1.10 | 1.37 ± 0.45 | 3.07 ± 0.82 | 5.10 ± 1.35 |
| Krishna | 20.68 | 0.75 ± 0.32 | 1.17 ± 0.50 | 1.17 ± 0.51 | 0.78 ± 0.33 | 1.61 ± 0.54 | 2.09 ± 0.65 | 0.76 ± 0.35 | 1.84 ± 0.62 | 3.01 ± 0.89 | 0.88 ± 0.38 | 2.27 ± 0.68 | 3.77 ± 1.07 |
| Mahanadi | 20.78 | 0.69 ± 0.22 | 1.12 ± 0.40 | 1.14 ± 0.41 | 0.67 ± 0.33 | 1.50 ± 0.46 | 1.95 ± 0.53 | 0.64 ± 0.37 | 1.75 ± 0.52 | 3.01 ± 0.69 | 0.75 ± 0.39 | 2.21 ± 0.55 | 3.91 ± 0.81 |
| Mahi | 19.85 | 0.93 ± 0.47 | 1.32 ± 0.60 | 1.36 ± 0.67 | 0.92 ± 0.46 | 1.77 ± 0.70 | 2.34 ± 0.92 | 0.84 ± 0.44 | 2.06 ± 0.79 | 3.46 ± 1.13 | 1.06 ± 0.53 | 2.59 ± 0.93 | 4.39 ± 1.46 |
| Narmada | 19.50 | 0.85 ± 0.41 | 1.29 ± 0.56 | 1.32 ± 0.59 | 0.83 ± 0.43 | 1.72 ± 0.64 | 2.27 ± 0.81 | 0.75 ± 0.43 | 1.99 ± 0.74 | 3.42 ± 1.04 | 0.96 ± 0.48 | 2.54 ± 0.80 | 4.41 ± 1.27 |
| North East Coast | 21.70 | 0.61 ± 0.19 | 1.04 ± 0.39 | 1.04 ± 0.38 | 0.62 ± 0.29 | 1.37 ± 0.43 | 1.78 ± 0.48 | 0.61 ± 0.32 | 1.61 ± 0.48 | 2.74 ± 0.63 | 0.68 ± 0.34 | 2.01 ± 0.51 | 3.50 ± 0.75 |
| Pennar | 21.92 | 0.64 ± 0.28 | 1.01 ± 0.45 | 1.01 ± 0.45 | 0.66 ± 0.29 | 1.40 ± 0.48 | 1.79 ± 0.58 | 0.68 ± 0.33 | 1.65 ± 0.53 | 2.67 ± 0.77 | 0.76 ± 0.33 | 2.02 ± 0.57 | 3.35 ± 0.96 |
| Sabarmati | 20.25 | 1.08 ± 0.40 | 1.49 ± 0.55 | 1.51 ± 0.65 | 1.07 ± 0.41 | 2.02 ± 0.65 | 2.62 ± 0.85 | 1.05 ± 0.43 | 2.33 ± 0.78 | 3.82 ± 1.11 | 1.23 ± 0.51 | 2.89 ± 0.88 | 4.76 ± 1.38 |
| South Coast | 22.95 | 0.76 ± 0.27 | 1.09 ± 0.45 | 1.09 ± 0.45 | 0.77 ± 0.30 | 1.49 ± 0.51 | 1.89 ± 0.64 | 0.80 ± 0.31 | 1.81 ± 0.54 | 2.88 ± 0.81 | 0.90 ± 0.33 | 2.17 ± 0.61 | 3.56 ± 0.93 |
| Subarnarekha | 20.67 | 0.63 ± 0.18 | 1.05 ± 0.29 | 1.09 ± 0.31 | 0.58 ± 0.32 | 1.35 ± 0.37 | 1.80 ± 0.43 | 0.55 ± 0.36 | 1.51 ± 0.51 | 2.67 ± 0.65 | 0.64 ± 0.42 | 2.00 ± 0.45 | 3.58 ± 0.67 |
| Tapi | 20.77 | 0.91 ± 0.44 | 1.37 ± 0.57 | 1.39 ± 0.61 | 0.91 ± 0.43 | 1.83 ± 0.67 | 2.42 ± 0.84 | 0.85 ± 0.42 | 2.15 ± 0.77 | 3.57 ± 1.10 | 1.07 ± 0.48 | 2.69 ± 0.85 | 4.52 ± 1.30 |
| West Coast | 21.44 | 1.01 ± 0.40 | 1.47 ± 0.61 | 1.49 ± 0.64 | 1.04 ± 0.40 | 2.00 ± 0.64 | 2.59 ± 0.75 | 1.04 ± 0.37 | 2.29 ± 0.67 | 3.64 ± 0.93 | 1.18 ± 0.44 | 2.79 ± 0.75 | 4.46 ± 1.06 |